\documentclass[final,,twoside]{IEEEtranTCOM}
\normalsize
\ifCLASSINFOpdf
\else
\fi

\usepackage{amsthm}
\usepackage{amsmath}
\usepackage{amssymb}
\usepackage{graphicx}
\usepackage{esint}
\usepackage{amsfonts}
\usepackage{cite}
\usepackage{balance}
\usepackage{caption}
\usepackage{subcaption}
\usepackage{epstopdf}
\usepackage{color}
\usepackage{tikz}
\usepackage{longtable}

\makeatletter

\theoremstyle{plain}
\newtheorem{thm}{\protect\theoremname}

\makeatother

\providecommand{\theoremname}{Theorem}

\theoremstyle{plain}
\newtheorem{lem}{\protect\lemmaname}

\makeatother

\providecommand{\lemmaname}{Lemma}

\DeclareMathOperator{\erf}{erf}

\begin{document}
\title{ Cascaded Composite Turbulence and Misalignment:  Statistical Characterization and Applications to Reconfigurable Intelligent Surface-Empowered Wireless~Systems }
\author{\vspace{-0.2cm} 
Alexandros-Apostolos A. Boulogeorgos, Senior Member, IEEE, Nestor Chatzidiamantis, Member, IEEE,  \\ Harilaos G. Sandalidis, Angeliki Alexiou, Member, IEEE, and Marco Di Renzo, Fellow, IEEE  

\thanks{A.-A. A. Boulogeorgos and A. Alexiou are with the of Digital Systems, University of Piraeus Piraeus 18534 Greece (e-mails: al.boulogeorgos@ieee.org, alexiou@unipi.gr).
}
\thanks{ N. Chatzidiamantis is with the Department of Electrical and Computer Engineering, Aristotle University of Thessaloniki, 54124 Thessaloniki, (e-mail: nestoras@auth.gr).
}
\thanks{ H. G. Sandalidis is with the Department of Computer Science and Biomedical Informatics, University of Thessaly, 35131 Lamia, Greece, (e-mail: sandalidis@dib.uth.gr).
}

\thanks{M. Di Renzo is with Universit\'e Paris-Saclay, CNRS, CentraleSup\'elec, Laboratoire des Signaux et Syst\`emes, 3 Rue Joliot-Curie, 91192 Gif-sur-Yvette, France. (marco.di-renzo@universite-paris-saclay.fr)} 

\thanks{This work has received funding from the European Unions Horizon 2020
research and innovation programme under grant agreement No. 871464
(ARIADNE).}

\thanks{A part of this work was submitted to IEEE International Conference on Communications (ICC) 2021.}
}
\maketitle	
\vspace{-0.4cm}
\begin{abstract}
 Reconfigurable intelligent surfaces (RISs) empowered high-frequency (HF) wireless systems are expected to become the supporting pillar for several reliability and data-rate hungry applications. Such systems are, however, sensitive to misalignment and atmospheric phenomena including turbulence. Most of the existing studies on the performance assessment of RIS-empowered wireless systems ignore the impact of the aforementioned phenomena. Motivated by this, the current contribution presents a theoretical framework for {statistically characterizing cascaded composite turbulence and misalignment channels.} More specifically, we {present the probability density and cumulative distribution functions for} the cascaded composite turbulence and misalignment channels. Building upon the derived analytical expressions {and in order to demonstrate the applicability and importance of the extracted framework in different use case cases of interest}, we present novel closed-form formulas that quantify the joint impact of turbulence and misalignment on the outage performance for two scenarios, namely cascaded multi-RIS-empowered free space optics (FSO) and terahertz (THz) wireless systems. {For the aforementioned scenarios, the diversity order is extracted.} In addition, we provide an insightful outage probability upper-bound for a third scenario that considers parallel multi-RIS-empowered FSO systems. Our results highlight the importance of accurately modeling both turbulence and misalignment when assessing the performance of such~systems.        
\end{abstract}
\vspace{-0.2cm}

\vspace{-0.4cm}
\section{Introduction}\label{S:Intro}
\vspace{-0.2cm}

By envisioning unprecedented performance requirements in terms of data-rates, reliability, availability, and security, the sixth generation (6G) wireless era comes with the promise to become the pillar of several ``killer-applications'', including but not limited to extended reality, high-speed kiosks, and wireless backhauling~\cite{latva2020key,Bariah2020,
	WP:Wireless_Thz_system_architecture_for_networks_beyond_5G,
	Dang2020}. To achieve these goals, two technology enablers are identified and expected to be exploited, namely high-frequency (HF) communications, in the sub-terahertz, terahertz  (THz)~\cite{Boulogeorgos2018,Zhang2019,Akyildiz2020,Boulogeorgos2020} and optical bands~\cite{Chi2020,Porambage2021,Arfaoui2021}, and  reconfigurable intelligent surfaces (RISs)~\cite{Renzo2019,Renzo2020,Tsilipakos2020,Boulogeorgos2020a}. THz and optical wireless systems can support aggregated data rates that may exceed $1\text{ }\mathrm{Tb/s}$~\cite{Boulogeorgos2018,Koenig2013,Parca2013,Esmail2016,Mazurczyk2020}; {thus, as envisioned, theoretically investigated and experimentally validated in~\cite{Jastrow2010,Kallfass2015,Castro2020,Boulogeorgos2021a,WP:Wireless_Thz_system_architecture_for_networks_beyond_5G,C:Performance_Evaluation_of_THz_Wireless_Systems_Operating_in_475_400_GHz_band,Elschner2020,Rodrigues2020,Elzanaty2020,Alzenad2018,Esmail2017,McDonald2021}, they can be used as wireless fiber extenders. Indicatively, in~\cite{Elschner2020}, a $1\text{ }\mathrm{km}$ wireless fiber extender was documented that operates at $300\text{ }\mathrm{GHz}$, while, in~\cite{Rodrigues2020}, a $500\text{ }\mathrm{m}$ THz demonstrator was reported. In~\cite{Esmail2017}, a $100\text{ }\mathrm{m}$, {FSO} demonstrator was presented. In~\cite{McDonald2021}, finally, an $800\text{ }\mathrm{m}$ outdoor point-to-point optical wireless link was demonstrated.}
	
	{Despite the paramount importance that HF systems can play as wireless fiber extenders} due to their sensitivity to blockages, {they are unable to} guarantee neither high-reliability nor continuous availability~\cite{Zhang2018,C:UserAssociationInUltraDenseTHzNetworks,
	Djordjevic2016,Kaushal2017}.  On the other hand, RISs are programmable metasurfaces, which are capable of altering the electromagnetic characteristics of the propagation medium, thus transforming conventional communication environments into smart platforms~\cite{Renzo2019,Liaskos2020,Renzo2020a,Tang2020}. In more detail, two of the most commonly examined RIS functionalities are blockage avoidance by creating alternative paths through beam-steering~\cite{Taghvaee2020,Zhou2021,Pan2020}, which ensures uninterrupted connectivity between the source ($S$) and the final destination ($D$), and beam focusing in order to extend the system transmission distance~\cite{Lee2014,Li2020}.    

Motivated by these considerations, a great amount of effort has been very recently put on combining the aforementioned technologies and designing RISs that operate {either} in the THz {or the} optical band~\cite{Cai2018,Shabanpour2020,Venkatesh2020,Amin2021,Tal2020,Ojaroudi2021,Manjappa2018}. {In~\cite{Cai2018} and~\cite{Shabanpour2020}, the authors presented a vanadium dioxide ($\text{VO}_2$)-based multi-functional RIS capable of manipulating THz waves, whereas a large-scale RIS that employs arrays of complementary metal-oxide-semiconductor (CMOS)-based chip tiles and operates at $0.3\text{ }\mathrm{THz}$ was documented in~\cite{Venkatesh2020}.} In~\cite{Amin2021}, a THz graphene plasmonic metasurface was reported, while a broadband nonlinear plasmonic metasurface THz emitter operating in the $\mathrm{THz}$ band was demonstrated in~\cite{Tal2020}. Similarly, a graphene-based RIS structure that provides beam steering and focusing capabilities at frequencies around $4.35\text{ }\mathrm{THz}$ was discussed in~\cite{Ojaroudi2021}. {In~\cite{Manjappa2018}, a micro-electro-mechanical (MEM)-based THz RIS was demonstrated.}

In addition, optical RISs have attracted a considerable attention~\cite{Brandl2013,Decker2015,Khorasaninejad2016,Wang2018,Zhang2020,Wang2020,Wang2017}. In~\cite{Brandl2013}, a MEM was employed as an optical RIS in order to provide beam-steering capabilities in an optical wireless communication (OWC) system, while all-dielectric metasurfaces capable of manipulating near infrared waves were reported in~\cite{Decker2015}. The authors of ~\cite{Khorasaninejad2016} presented metalens-based metasurfaces that enable diffraction-limited focusing at wavelengths of $405$, $532$, and $660 \text{ }\mathrm{nm}$, whereas an achromatic gallium nitride (GaN)-based optical RIS that operates in the entire visible region was reported in~\cite{Wang2018}. In~\cite{Zhang2020}, an optical RIS, which is capable of controlling the angular dispersions, enabling functionalities that range from perfect mirroring to  angular multiplexing, was documented.  In~\cite{Wang2020}, in addition, two types of optical RISs were presented based on MEMs and phase array based~technologies. {Finally, in~\cite{Wang2017}, an achromatic metasurface capable of eliminating the chromatic aberration over a continuous region from $1200$ to $1680\text{ }\mathrm{nm}$ for circularly-polarized incidence light in a reflection scheme was reported.} 

From the performance analysis, system design, and optimization points of view, the error performance of RIS-empowered THz wireless satellite systems in the presence of antenna misalignment were studied in~\cite{Tekbiyik2020}. {In~\cite{Chapala2021}, the outage, error, and capacity performance of a single-RIS-empowered THz wireless system in the presence of $\alpha-\mu$ fading and antenna misalignment was analyzed. In~\cite{Papasotiriou2020}, a capacity evaluation of RIS-empowered sub-THz wireless systems was conducted.} In~\cite{Du2020}, the authors assessed the joint impact of hardware imperfections and antenna misalignment on RIS-empowered indoor THz wireless systems, whereas the coverage performance of RIS-empowered THz wireless systems were quantified in~\cite{Boulogeorgos2021}. In~\cite{Huang2021}, the authors introduced a multi-RIS-empowered THz hybrid beamforming architecture and formulated the design problem of the digital and analog beamforming matrices, assuming that all the intermediate channels, between the transmitter and the receiver, experience neither fading nor misalignment. In~\cite{Wan2021}, in addition, a holographic RIS-empowered THz massive multiple-input multiple-output (MIMO) system accompanied by a low-overhead closed-loop channel estimation scheme was presented.  In~\cite{Lu2020} the authors presented a sum-rate maximization framework for RIS-empowered THz wireless systems, by assuming that the transmitter-RIS and RIS-destination channels have a Rician distribution. {In~\cite{Abuzainab2021}, the authors studied the problem of proactive handoff and beam selection in THz drone communication networks assisted with RIS. Similarly, in~\cite{Pan2021}, the joint optimization problem of drone's trajectory, the phase shift of RIS, the allocation of THz sub-bands, and the power control were investigated aiming to maximize the minimum average achievable rate of all users. In~\cite{Pan2021a}, finally, a simultaneous terahertz (THz) information and power transfer system was introduced and the problem of maximizing the information users' data rate while ensuring the energy users' and RIS's power harvesting requirements was formulated and solved.}
It is worth mentioning that all the aforementioned works assume that the RISs have horizontal and vertical meta-atom periodicity of the order of half  of the~wavelength.

The performance of RIS-empowered OWC systems was analyzed in~\cite{Abumarshoud2021,Wang2020a,Najafi2020,Ndjiongue2021}. In~\cite{Abumarshoud2021}, the received signal-to-noise-ratio (SNR) in RIS-empowered visible light communication systems were evaluated. In~\cite{Wang2020a}, the authors studied the performance of a single-RIS-empowered OWC system in the presence of random obstacles and pointing errors, assuming that the impact of turbulence can be neglected. In~\cite{Najafi2020}, the authors considered a single-RIS free-space-optics (FSO) system and quantified its outage performance by devising an equivalent mirror-assisted FSO system that generates a reflected electric field on a mirror, which is identical to that on the source. Although this approach seems very promising, it cannot capture the statistics of the two cascaded links. Aspired by this, a closed-form expression for the outage probability (OP) of a single-RIS-empowered FSO system, in which both links experience different levels of turbulence and misalignment, was derived in~\cite{Ndjiongue2021}.        

Recently, the concept of using multiple-RISs to provide uninterrupted connectivity between pairs of transmitters and receivers was reported in~\cite{Liaskos2018}. {Multi-RIS can enable a number of novel functionalities including, but not limited to blockage avoidance, routing, coverage expansion, and beam splitting.} Despite the importance and capabilities that this concept can offer, to the best of the authors' knowledge, a theoretical framework for analyzing the achievable performance of such systems has not been formulated yet. This is due to the lack of a statistical model for the cascaded composite turbulence and misalignment fading channels. Motivated by this, in this paper, we introduce an analytical framework that covers the aforementioned gap as well as its applications in RIS-empowered wireless systems. In particular, the technical contribution of this work is as follows:
\begin{itemize}
	\item We present closed-form expressions for the probability density function (PDF) and cumulative distribution function (CDF) for cascaded channels that experience different levels of turbulence.
	\item We derive a novel analytical expression for the PDF of cascaded channels that experience misalignment fading.
	\item Building upon the aforementioned expressions, we statistically characterize cascaded channels that experience  different levels of turbulence and misalignment fading, in terms of PDF and CDF.
	\item We apply the obtained analytical frameworks for  quantifying the outage performance of three relevant application scenarios, namely: i) a cascaded multi-RIS empowered FSO system, ii) a multi-RIS empowered FSO systems, and iii) a cascaded multi-RIS empowered THz wireless system. More precisely, in the first scenario, we introduce an accurate closed-form expression for the system OP that takes into account both the characteristics of the transmitter, receiver and propagation channel. For the second scenario, we report an OP upper-bound, and for the third scenario, we provide a closed-form expression for the OP that accounts for the impact of transceivers hardware~imperfections.  
	\item {We provide the diversity order of the cascaded multi-RIS empowered FSO and THz systems.}     
\end{itemize}   

The rest of this contribution is structured as follows: The statistical characterization of cascaded wireless channels that experience turbulence and/or misalignment fading is presented in~Section~\ref{S:SC}. Some applications of the analytical framework are discussed in~Section~\ref{S:App}, while respective Monte Carlo simulations that verify the theoretical framework, accompanied by insightful discussions, are illustrated in~Section~\ref{S:Results}. Finally, a summary of our main findings and concluding remarks are provided in~Section~\ref{S:Conclusions}.                 

\subsubsection*{Notations} 
The absolute value, exponential and natural logarithm functions are respectively denoted by $|\cdot|$,  $\exp\left(\cdot\right)$, and $\ln\left(\cdot\right)$.
  $\sqrt{x}$ and $\prod_{l=1}^{L}x_l$ respectively return the square root of $x$,  and the product of $x_1\,x_2\,\cdots\, x_L$.  
 $\Pr\left(\mathcal{A}\right)$ denotes the probability for the event $\mathcal{A}$ to be valid. 
The modified Bessel function of the second kind of order $n$ is denoted as~$\mathrm{K}_n(\cdot)$~\cite[eq. (8.407/1)]{B:Gra_Ryz_Book}. 
The  Gamma~\cite[eq. (8.310)]{B:Gra_Ryz_Book} function is  denoted by  $\Gamma\left(\cdot\right)$, and the error-function is represented by $\erf\left(\cdot\right)$~\cite[eq. (8.250/1)]{B:Gra_Ryz_Book}. {The generalized hypergeometric function is denoted by $\,_p\mathrm{F}_q\left(a;b;z\right)$.} Finally,  $G_{p, q}^{m, n}\left(x\left| \begin{array}{c} a_1, a_2, \cdots, a_{p} \\ b_{1}, b_2, \cdots, b_q\end{array}\right.\right)$  stands for the Meijer G-function~\cite[eq. (9.301)]{B:Gra_Ryz_Book}. {Tables~\ref{T:Parameters_definition} and~\ref{T:RV_definition} provide the definitions of all the symbols that are employed in this paper.}

\begin{table}
\begin{center}
	\caption{{Parameters and performance indicators definition.}}
	\label{T:Parameters_definition}
	\begin{tabular}{|c| p{7.4cm}|}
		\hline
		\multicolumn{2}{|c|}{\textbf{Parameters}} \\
		\hline \hline
		$p$ & Atmospheric pressure\\
		$T$ & Atmospheric temperature\\
		$P_s$ & Average power constraint of the FSO signal\\
		$B$ & Bandwidth\\
		$k_B$ & Boltzman's constant\\
		$A_Q$ & Derivative of the real-part of the refraction index with respect to the relative humidity\\
		$A_T$ & Derivative of the real-part of the refraction index with respect to the temperature\\
		$g_i$ & Deterministic path-gain coefficient of the $i-$th THz link\\
		$d_i$ & Distance of the $i-$th link\\
		$g_{f,i}$ & Free-space path-gain coefficient of the $i-$th THz link\\
		$\tau(f, d)$ & Molecular absorption gain coefficient\\
		$\kappa(f)$ & Molecular absorption coefficient\\
		$N$ & No. of Gamma-Gamma (GG) distributed random variables (RVs)\\
		$L$ & No. of misaligned links\\
		$f$ & Operation frequency\\
		$G_d$ & Reception antenna gain\\
		$\phi$ & Relative humidity\\ 
		$p_w$ & Saturated water vapor partial pressure\\
		$C_T$ & Structure factor\\
		$G_s$ & Transmission antenna gain\\
		$s_i$ & Transmission signal of the $i-$th $S$ at the parallel multi-RIS FSO \\
		$\mu_w$& Volume mixing ratio of the water vapor\\
		$\eta$ & Photodetector (PD) responsivity\\ 
		$c$ & Speed of light \\
		$\lambda$ & Wavelength of the optical carrier\\ 
		$\alpha$ & Weather dependent attenuation coefficient \\
		$G_i$ & $i-$th FSO link atmospheric gain \\
		$R_i$ & $i-$th RIS reflection coefficient\\
		$\rho_i$ & $i-$th optical RIS reflection efficiency \\
		$s$ & Intensity of the FSO transmitted signal \\ 
		$x$ & Transmission signal at the RIS-empowered THz wireless system\\
		$\rho_s$ & Transmission SNR multiplied by the deterministic path-gain of the multi-RIS FSO system\\
		$r_{\mathrm{th}}$ & SNR threshold at the multi-RIS FSO system\\
		$\rho_{\mathrm{th}}$ & SNR threshold at the parallel multi-RIS FSO system\\
		$\alpha_i$, $\beta_i$ & Shaping parameters of $r_i$\\
		$\Omega_i$ & Mean of $r_i$\\
		$\xi_i$, $A_{o,i}$ & Distribution parameters of $l_i$\\ 
		$\sigma_n^2$ & Variance of the additive white Gaussian shot noise\\
		$C_{n}^2$ & Refraction index parameters \\
		$\sigma_{R_i}^2$ & Rytov variance of the $i-$th FSO link\\
		$b_i$ & Radius of the circular aperture at the $i-$th RIS or D plane\\ 
		$w_{d_i}$ & Beam waste at the $i-$th RIS or D plane\\
		$w_{\mathrm{eq},i}$ & Equivalent beam radius at the $i-$th RIS or $D$ plane\\
		$\sigma_{s,i}^2$ & Pointing error displacement (jitter) variance\\
		$N_o$ & Noise variance \\
		$\sigma_s^2$ & Variance of $\eta_s$\\
		$\sigma_d^2$ & Variance of $\eta_d$\\
		$\kappa_s$ & Error vector magnitude of the $S$ transmitter \\
		$\kappa_r$ & Error vector magnitude of the $D$ receiver\\
		\hline
		\multicolumn{2}{|c|}{\textbf{Performance indicators}} \\
		\hline \hline
		$P_{o}^{\text{FSO}}$ & Multi-RIS FSO system OP\\
		$P_{o}^{\text{par}}$ & Parallel multi-RIS FSO system OP\\
		$P_{o}^{\text{THz}}$ & RIS-empowered THz wirelss system OP\\
		\hline
	\end{tabular}
\end{center}
\end{table}
\begin{table}
	\begin{center}
		\caption{{Random variables~(RVs) definition.}}
		\label{T:RV_definition}
		\begin{tabular}{|c| p{7.4cm}|}
			\hline
			$r_i$ & $i-$th GG distributed RV\\
			$l_i$ & Misalignment fading coefficient of the $i-$th misaligned link\\  
			$Z_1$ & Product of independent distributed GG RVs\\
			$Z_2$ & Product of $l_i$, $i=1,\cdots, L$ RVs\\
			$A$ & Cascaded multi-RIS empowered FSO end-to-end (e2e) channel\\
			$h_i$ & Turbulence coefficient of the $i-$th FSO link\\
			$h_{p,i}$ & Misalignment fading coefficient of the $i-$th FSO~link\\
			$B_i$ & E2e channel coefficient of the $i-$th FSO link\\ 
			$h_{1,i}$ & $i-$th $S$-RIS link turbulence coefficient \\
			$h_{2,i}$ & $i-$th RIS-$D$ link turbulence coefficient \\
			$h_{p1,i}$ & $i-$th $S$-RIS link misalignment fading coefficient\\
			$h_{p2,i}$ &  $i-$th RIS-$D$ link misalignment fading coefficient\\
			$n$ & Additive white Gaussian shot noise \\
			$A_t$ & THz e2e channel coefficient \\
			$h_{p,i}^{\text{THz}}$ & Misalignment fading coefficient of the $i-$th THz~link\\
			$h_{t,i}^{\text{THz}}$ & Turbulence coefficient of the $i-$th THz~link\\
			$w$ & AWGN at the $D$ of the RIS-empowered THz wireless system\\ 
			$r$ & Received signal at the multi-RIS FSO system\\
			$r_p$ & Received signal at the parallel multi-RIS FSO system\\
			$\eta_{s}$ & $S$ distortion noise at the RIS-empowered THz wireless system\\
			$\eta_{d}$ & $D$ distortion noise at the RIS-empowered THz wireless system\\
			$y$ & Received signal at $D$ at the RIS-empowered THz wireless system\\
			$\rho_{\mathrm{FSO}}$ & Instantaneous SNR at D of the  multi-RIS FSO system\\
			$\rho_{\mathrm{par}}$ & Instantaneous SNR at D of the parallel multi-RIS FSO system\\
			$\gamma$ & Signal-to-distortion-plus-noise-ratio (SDNR)\\
			\hline
	\end{tabular}
\end{center}
\end{table}

\vspace*{-0.6cm}
\section{Statistical Characterization of Product~Channels}\label{S:SC}
\vspace{-0.2cm}

\setlength{\abovedisplayskip}{1pt} \setlength{\abovedisplayshortskip}{1pt}
\setlength{\belowdisplayskip}{1pt} \setlength{\belowdisplayshortskip}{1pt}
Let $Z_{1}$ be the product of  $N\geq 1$ independently distributed Gamma-Gamma (GG) random variables (RVs), i.e.,
\begin{align}
	Z_{1} = \prod_{i=1}^{N} r_{i},
	\label{Eq:Z_1}
\end{align}   
where $r_{i}$, with $i\in[1, N]$, is the $i-$th distributed GG RV. The PDF of $r_{i}$ can be written as follows~\cite{Denic2008,Benkhelifa2013}
\begin{align}
	f_{r_{i}}(x) &= \frac{2}{\Gamma\left(\alpha_i\right) \Gamma\left(\beta_i\right)} \left(\frac{\alpha_i \beta_i}{\Omega_i}\right)^{\frac{\alpha_i + \beta_i}{2}}  x^{\frac{\alpha_i + \beta_i}{2}-1}
	\nonumber \\
	\times & \mathrm{K}_{\alpha_i-\beta_i}\left(\alpha_i-\beta_i, 2 \sqrt{\frac{\alpha_i \beta_i x}{\Omega_i}} \right),
	\label{Eq:f_r_i}
\end{align}
where $\alpha_i \geq 0$ and $\beta_{i} \geq 0$ are the shaping parameters of the $r_i$ distribution and $\Omega_i$ is the corresponding mean, i.e., $\Omega_i = \mathbb{E}_{r_i}\left[x\right]$. Of note, the distribution in~\eqref{Eq:f_r_i} is generic, since, for different combinations of $\alpha_i$ and $\beta_i$, it returns various models that are usually used in communication systems. For $\alpha_i\to\infty$, it approximates the Gamma distribution, which have been widely used to model the fading  of radio frequency systems.  For $\alpha_i = 1$, it returns the $K$ distribution that is suitable for wireless systems with strong line-of-sights components, such as in HF systems. For the special case of $\alpha_i=\beta_i=1$, finally, it reduces to the PDF of the double Rayleigh~distribution. 

The following theorem provides closed-form expressions for the evaluation of the PDF and the CDF of $Z_{1}$. 
\begin{thm}
	The PDF of $Z_1$ can be evaluated~as
	\begin{align}
		f_{Z_{1}}(x)\hspace{-0.1cm} =\hspace{-0.1cm} \frac{\mathrm{G}_{0, 2N}^{2N, 0}\hspace{-0.1cm}\left[\hspace{-0.1cm} x \prod_{i=1}^{N} \frac{\alpha_i \beta_i}{\Omega_i} \Big| \alpha_1, \beta_1, \alpha_2, \beta_2 \cdots, \alpha_N, \beta_N \big. \right]}{x\prod_{i=1}^{N} \Gamma\left(\alpha_i\right) \Gamma\left(\beta_i\right)},
		\label{Eq:Z_1} 
	\end{align}\vspace{-0.1cm}
	whereas, its CDF can be obtained~as
	\begin{align}
		&F_{Z_{1}}(x) = \frac{1}{\prod_{i=1}^{N} \Gamma\left(\alpha_i\right) \Gamma\left(\beta_i\right)}
		\nonumber \\ & \times
		\mathrm{G}_{1, 2N+1}^{2N, 1}\left[ x \prod_{i=1}^{N} \frac{\alpha_i \beta_i}{\Omega_i} \Big| \begin{array}{c} 1 \\ \alpha_1, \beta_1, \alpha_2, \beta_2 \cdots, \alpha_N, \beta_N, 0 \end{array} \big. \right].
		\label{Eq:F_Z_1}
	\end{align} 
\end{thm}   
\begin{IEEEproof}
	{The product of gamma-gamma RVs is a special case of the product of generalized-gamma RVs derived in~\cite[eqs. (2) and (9)]{A:Ratio_of_products_of_alpha_mu_variates}. By setting the suitable parameter in~\cite[eqs. (2) and (9)]{A:Ratio_of_products_of_alpha_mu_variates} and after some algebraic manipulations, we extract ~\eqref{Eq:Z_1} and~\eqref{Eq:F_Z_1}. This concludes the~proof.}   
\end{IEEEproof}

Let $Z_{2}$ be the product of $L$, $L\geq 1$, independent RVs, i.e.,
\begin{align}
	Z_{2} = \prod_{i=1}^{L} l_i,
	\label{Eq:Z_2}
\end{align}
where $l_i$ is the $i-$th RV and its PDF can be expressed as 
\begin{align}
	f_{l_i}(x) = \frac{\xi_i}{A_{o,i}^{\xi_i}} x^{\xi_i-1},\text{ with } 0\leq x \leq A_{o,i},
	\label{Eq:f_l_i}
\end{align}
with $\xi_i$ and $A_{o,i}$ being the distribution parameters of $l_i$. The following theorem provides a closed-form expression for the PDF of $Z_{2}$.  
\begin{thm}
	The PDF of $Z_2$ can be expressed~as
	\begin{align}
		f_{Z_{2}}(x) &= \frac{1}{(L-1)!}
		\frac{\prod_{i=1}^{L}\xi_{i}}{\prod_{i=1}^{L}A_{o,i}^{\xi_i}} x^{\xi_L-1}
		\nonumber \\ & \times \left(\ln\left(\frac{\prod_{i=1}^{L}A_{o,i}}{x}\right)\right)^{L-1}
		\label{Eq:f_Z_2}
	\end{align}
 with  $0\leq x \leq \prod_{i=1}^{L}A_{o,i}$.
\end{thm}
\begin{IEEEproof}
	For brevity, the proof of Theorem 2 is given in Appendix B.    
\end{IEEEproof}

For the special case in which $A_{o,1}=A_{o,2}=\cdots=A_{o,M}=A_o$ and $\xi_1=\xi_2=\cdots=\xi_M=\xi$,~\eqref{Eq:f_Z_2} can be simplified~to
\begin{align}
	f_{Z_2}^{\text{sc}}(x) = \frac{1}{(L-1)!} \left(\frac{\xi}{A_o^{\xi}}\right)^{L} x^{\xi-1} \left(\ln\left(\frac{A_o^{L}}{x}\right)\right)^{L-1},
\end{align} 
with  $0\leq x \leq A_o^{L}$.

\begin{thm}
	The PDF and CDF of 
	\begin{align}
		Z= Z_1 \, Z_2,
		\label{Eq:Z}
	\end{align}
	can be formulated as shown in~\eqref{Eq:f_Z_final} and~\eqref{Eq:F_Z_final}, given at the top of the next page.
	\begin{figure*}
		\begin{align}
			f_{Z}(x) &= 
			\frac{\prod_{i=1}^{L}\xi_i}{\prod_{i=1}^N\Gamma\left(\alpha_i\right) \Gamma\left(\beta_i\right)} 
			x^{-1}
			\mathrm{G}_{ L,2N+L}^{ 2N+L,0}\left( \frac{1}{\prod_{i=1}^{N}\frac{\Omega_i}{\alpha_i\beta_i}\prod_{i=1}^{L}A_{o,i}} x
			\left| 
			\begin{array}{c}
				1+\xi_1, 1+\xi_2, \cdots, 1+\xi_L \\
				\alpha_{1},\cdots,\alpha_{N}, \beta_{1}, \cdots, \beta_{N}, \xi_1, \cdots, \xi_L
			\end{array}\right. \right)
			\label{Eq:f_Z_final}
		\end{align}
		\hrulefill
	\end{figure*}
\begin{figure*}
	\begin{align}
		F_{Z}(x) = \frac{\prod_{i=1}^{L}\xi_i}{\prod_{i=1}^N\Gamma\left(\alpha_i\right) \Gamma\left(\beta_i\right)} \mathrm{G}^{2N+L,1}_{L+1,2N+L+1}\left(\frac{1}{\prod_{i=1}^{N}\frac{\Omega_i}{\alpha_i\beta_i}\prod_{i=1}^{L}A_{o,i}} x
		\left|\begin{array}{c}1, \xi_1+1, \cdots, \xi_L+1 \\  \alpha_1, \cdots, \alpha_N, \beta_1, \cdots, \beta_N, \xi_1, \cdots, \xi_L, 0  \end{array}\right.\right)
		\label{Eq:F_Z_final}
	\end{align}
	\hrulefill
\end{figure*}
\end{thm}
\begin{IEEEproof}
	For brevity, the proof of Theorem 3 is given in Appendix C.    
\end{IEEEproof}
{It should be noted that the expressions of \eqref{Eq:f_Z_final} and~\eqref{Eq:F_Z_final} can be written in terms of more familiar hypergeometric functions according to  \cite[Eq. (9.303)]{B:Gra_Ryz_Book}. In more detail, for the special, but very realistic, case in which no two elements of the tuple $\mathcal{B}=\left[\alpha_1, \cdots, \alpha_N, \beta_1, \cdots, \beta_N, \xi_1, \cdots, \xi_L, 0\right]$ differ by an integer,~\eqref{Eq:F_Z_final} can be rewritten~as~in~\eqref{Eq:FZalternative}, given at the top of the next~page. 
	\begin{figure*}
		\begin{align}
			F_{Z}(x) &= \frac{\prod_{i=1}^{L}\xi_i}{\prod_{i=1}^N\Gamma\left(\alpha_i\right) \Gamma\left(\beta_i\right)} 
			\sum_{i=1}^{2N+L} \frac{\prod_{j=1}^{2N+L}\Gamma\left(\mathcal{B}_j-\mathcal{B}_i\right)\Gamma\left(\mathcal{A}_i\right)}{\prod_{j=2N+L}^{2N+L+1}
				\Gamma\left(1+\mathcal{B}_i-\mathcal{B}_j\right)\prod_{j=2}^{L+1}\Gamma\left(\mathcal{A}_j-\mathcal{B}_i\right)} x^{\mathcal{B}_i}
			\nonumber \\ & \times 
			\,_{L+1}\mathrm{F}_{2N+L}\left(1+\mathcal{B}_i-\mathcal{A}_1,\cdots,1+\mathcal{B}_i-\mathcal{A}_{L+1}; 1+\mathcal{B}_i-\mathcal{B}_1, \cdots  1+\mathcal{B}_i-\mathcal{B}_{2N+L+1}; (-1)^{-2N} x
			\right) 
			\label{Eq:FZalternative}
		\end{align}
		\hrulefill
	\end{figure*}
	In~\eqref{Eq:FZalternative},~$\mathcal{A}_{i}$ with $i=1,\cdots, L+1$ refers to the $i-$th element of the tuple $\mathcal{A}=\left\{ 1, \xi_1+1, \cdots, \xi_L+1 \right\}$. 
	Similarly, $\mathcal{B}_i$ with $i=1,\cdots, 2N+L$ refers to the $i-$th of $\mathcal{B}$.
	According to~\cite{B:Prudnikov_v3,Wimp1964,Slater2008}, since Meijer-G can be expressed as a finite sum of hypergeometric functions, it is considered a closed-form expression. Moreover, notice that nowadays, the Meijer-G function can be directly computed in several software packages and programming languages, including, but not limited to, Mathematica, Maple, Matlab, Python,~C++.
}

{The following lemma return the limit of the CDF of $Z$ as $x\to 0$.
\begin{lem}
In the case in which no two elements of the tuple $\mathcal{B}$ differ by an integer, as $x$ tends to $0$, the limit of the CDF of $Z$ can be approximated~as
\begin{align}
	&F_{Z}^{0}(x) = \frac{\prod_{i=1}^{L}\xi_i}{\prod_{i=1}^N\Gamma\left(\alpha_i\right) \Gamma\left(\beta_i\right)} 
	\nonumber \\ & \times
	\sum_{i=1}^{2N+L} \frac{\prod_{j=1}^{2N+L}\Gamma\left(\mathcal{B}_j-\mathcal{B}_i\right)\Gamma\left(\mathcal{A}_i\right)}{\prod_{j=2N+L}^{2N+L+1}
		\Gamma\left(1+\mathcal{B}_i-\mathcal{B}_j\right)\prod_{j=2}^{L+1}\Gamma\left(\mathcal{A}_j-\mathcal{B}_i\right)} x^{\mathcal{B}_i}
	\label{Eq:FZalternative_to_0}
\end{align}
\end{lem}
\begin{IEEEproof}
As $x\to 0$, \begin{align}
	&\,_{L+1}\mathrm{F}_{2N+L}\left(1+\mathcal{B}_i-\mathcal{A}_1,\cdots,1+\mathcal{B}_i-\mathcal{A}_{L+1}; 
	\right. \nonumber \\ &\left.1+\mathcal{B}_i-\mathcal{B}_1, \cdots  1+\mathcal{B}_i-\mathcal{B}_{2N+L+1}; (-1)^{-2N} x
\right) \to 1
\label{Eq:F_limit}
\end{align}
By applying~\eqref{Eq:F_limit} into~\eqref{Eq:FZalternative}, we obtain~\eqref{Eq:FZalternative_to_0}. This concludes the~proof. 
\end{IEEEproof}}

\vspace{-0.4cm}
\section{Applications}\label{S:App}
\vspace{-0.2cm}

This section is focused on presenting some applications of the theoretical framework. In more detail, three scenarios are considered: i) cascaded multi-RIS empowered FSO; ii) parallel multi-RIS empowered FSO; and iii) cascaded multi-RIS empowered wireless THz systems. For each scenario, the corresponding systems models are first presented and then closed-form expressions or upper bounds for the OP are~computed.

\vspace{-0.6cm}
\subsection{Cascaded multi-RIS empowered FSO systems}\label{SS:FSO1}
\vspace{-0.2cm}

\subsubsection{System model}
\begin{figure}
	\centering
	\vspace{-0.3cm}
	\scalebox{.53}{
		
		\tikzset{every picture/.style={line width=0.75pt}} 
		
		\begin{tikzpicture}[x=0.75pt,y=0.75pt,yscale=-1,xscale=1]
			
			\draw  [fill={rgb, 255:red, 198; green, 216; blue, 238 }  ,fill opacity=1 ] (245,201) .. controls (242.78,178.91) and (256.66,161) .. (275.99,161) .. controls (295.32,161) and (312.78,178.91) .. (315,201) .. controls (317.22,223.09) and (303.34,241) .. (284.01,241) .. controls (264.68,241) and (247.22,223.09) .. (245,201) -- cycle ;
			\draw  [fill={rgb, 255:red, 173; green, 173; blue, 173 }  ,fill opacity=1 ] (256.09,168) -- (266.79,168) -- (267.89,179) -- (257.19,179) -- cycle ;
			\draw  [fill={rgb, 255:red, 173; green, 173; blue, 173 }  ,fill opacity=1 ] (271.22,167.99) -- (281.92,167.99) -- (283.02,178.99) -- (272.32,178.99) -- cycle ;
			\draw  [fill={rgb, 255:red, 173; green, 173; blue, 173 }  ,fill opacity=1 ] (287.62,168.99) -- (298.32,168.99) -- (299.42,179.99) -- (288.72,179.99) -- cycle ;
			\draw  [fill={rgb, 255:red, 173; green, 173; blue, 173 }  ,fill opacity=1 ] (248.22,182.99) -- (258.92,182.99) -- (260.02,193.99) -- (249.33,193.99) -- cycle ;
			\draw  [fill={rgb, 255:red, 173; green, 173; blue, 173 }  ,fill opacity=1 ] (262.22,182.99) -- (272.92,182.99) -- (274.02,193.99) -- (263.33,193.99) -- cycle ;
			\draw  [fill={rgb, 255:red, 173; green, 173; blue, 173 }  ,fill opacity=1 ] (277.22,182.99) -- (287.92,182.99) -- (289.02,193.99) -- (278.33,193.99) -- cycle ;
			\draw  [fill={rgb, 255:red, 173; green, 173; blue, 173 }  ,fill opacity=1 ] (293.62,183.99) -- (304.32,183.99) -- (305.42,194.99) -- (294.73,194.99) -- cycle ;
			\draw  [fill={rgb, 255:red, 173; green, 173; blue, 173 }  ,fill opacity=1 ] (249.83,196.99) -- (260.52,196.99) -- (261.63,207.99) -- (250.93,207.99) -- cycle ;
			\draw  [fill={rgb, 255:red, 173; green, 173; blue, 173 }  ,fill opacity=1 ] (263.83,196.99) -- (274.52,196.99) -- (275.63,207.99) -- (264.93,207.99) -- cycle ;
			\draw  [fill={rgb, 255:red, 173; green, 173; blue, 173 }  ,fill opacity=1 ] (278.83,196.99) -- (289.52,196.99) -- (290.63,207.99) -- (279.93,207.99) -- cycle ;
			\draw  [fill={rgb, 255:red, 173; green, 173; blue, 173 }  ,fill opacity=1 ] (295.23,197.99) -- (305.92,197.99) -- (307.03,208.99) -- (296.33,208.99) -- cycle ;
			\draw  [fill={rgb, 255:red, 173; green, 173; blue, 173 }  ,fill opacity=1 ] (251.83,211.99) -- (262.53,211.99) -- (263.63,222.99) -- (252.93,222.99) -- cycle ;
			\draw  [fill={rgb, 255:red, 173; green, 173; blue, 173 }  ,fill opacity=1 ] (265.83,211.99) -- (276.53,211.99) -- (277.63,222.99) -- (266.93,222.99) -- cycle ;
			\draw  [fill={rgb, 255:red, 173; green, 173; blue, 173 }  ,fill opacity=1 ] (280.83,211.99) -- (291.53,211.99) -- (292.63,222.99) -- (281.93,222.99) -- cycle ;
			\draw  [fill={rgb, 255:red, 173; green, 173; blue, 173 }  ,fill opacity=1 ] (297.23,212.99) -- (307.93,212.99) -- (309.03,223.99) -- (298.33,223.99) -- cycle ;
			\draw  [fill={rgb, 255:red, 173; green, 173; blue, 173 }  ,fill opacity=1 ] (267.31,226) -- (278.01,226) -- (279.11,237) -- (268.41,237) -- cycle ;
			\draw  [fill={rgb, 255:red, 173; green, 173; blue, 173 }  ,fill opacity=1 ] (282.43,225.99) -- (293.13,225.99) -- (294.24,236.99) -- (283.54,236.99) -- cycle ;
			\draw  [draw opacity=0][fill={rgb, 255:red, 201; green, 56; blue, 73 }  ,fill opacity=0.72 ] (290.57,185.96) .. controls (299.61,191.19) and (304.32,202.84) .. (301.09,211.97) .. controls (297.86,221.1) and (287.92,224.26) .. (278.88,219.03) .. controls (269.85,213.8) and (265.14,202.15) .. (268.36,193.02) .. controls (271.59,183.89) and (281.53,180.73) .. (290.57,185.96) -- cycle ;
			\draw  [fill={rgb, 255:red, 198; green, 216; blue, 238 }  ,fill opacity=1 ] (412,85) .. controls (418.63,62.91) and (439.67,45) .. (459,45) .. controls (478.33,45) and (488.63,62.91) .. (482,85) .. controls (475.37,107.09) and (454.33,125) .. (435,125) .. controls (415.67,125) and (405.37,107.09) .. (412,85) -- cycle ;
			\draw  [fill={rgb, 255:red, 173; green, 173; blue, 173 }  ,fill opacity=1 ] (436.3,52) -- (447,52) -- (443.7,63) -- (433,63) -- cycle ;
			\draw  [fill={rgb, 255:red, 173; green, 173; blue, 173 }  ,fill opacity=1 ] (451.43,51.99) -- (462.13,51.99) -- (458.83,62.99) -- (448.13,62.99) -- cycle ;
			\draw  [fill={rgb, 255:red, 173; green, 173; blue, 173 }  ,fill opacity=1 ] (467.43,52.99) -- (478.13,52.99) -- (474.83,63.99) -- (464.13,63.99) -- cycle ;
			\draw  [fill={rgb, 255:red, 173; green, 173; blue, 173 }  ,fill opacity=1 ] (422.43,66.99) -- (433.13,66.99) -- (429.83,77.99) -- (419.13,77.99) -- cycle ;
			\draw  [fill={rgb, 255:red, 173; green, 173; blue, 173 }  ,fill opacity=1 ] (436.43,66.99) -- (447.13,66.99) -- (443.83,77.99) -- (433.13,77.99) -- cycle ;
			\draw  [fill={rgb, 255:red, 173; green, 173; blue, 173 }  ,fill opacity=1 ] (451.43,66.99) -- (462.13,66.99) -- (458.83,77.99) -- (448.13,77.99) -- cycle ;
			\draw  [fill={rgb, 255:red, 173; green, 173; blue, 173 }  ,fill opacity=1 ] (467.43,67.99) -- (478.13,67.99) -- (474.83,78.99) -- (464.13,78.99) -- cycle ;
			\draw  [fill={rgb, 255:red, 173; green, 173; blue, 173 }  ,fill opacity=1 ] (418.43,80.99) -- (429.13,80.99) -- (425.83,91.99) -- (415.13,91.99) -- cycle ;
			\draw  [fill={rgb, 255:red, 173; green, 173; blue, 173 }  ,fill opacity=1 ] (432.43,80.99) -- (443.13,80.99) -- (439.83,91.99) -- (429.13,91.99) -- cycle ;
			\draw  [fill={rgb, 255:red, 173; green, 173; blue, 173 }  ,fill opacity=1 ] (447.43,80.99) -- (458.13,80.99) -- (454.83,91.99) -- (444.13,91.99) -- cycle ;
			\draw  [fill={rgb, 255:red, 173; green, 173; blue, 173 }  ,fill opacity=1 ] (463.43,81.99) -- (474.13,81.99) -- (470.83,92.99) -- (460.13,92.99) -- cycle ;
			\draw  [fill={rgb, 255:red, 173; green, 173; blue, 173 }  ,fill opacity=1 ] (414.43,95.99) -- (425.13,95.99) -- (421.83,106.99) -- (411.13,106.99) -- cycle ;
			\draw  [fill={rgb, 255:red, 173; green, 173; blue, 173 }  ,fill opacity=1 ] (428.43,95.99) -- (439.13,95.99) -- (435.83,106.99) -- (425.13,106.99) -- cycle ;
			\draw  [fill={rgb, 255:red, 173; green, 173; blue, 173 }  ,fill opacity=1 ] (443.43,95.99) -- (454.13,95.99) -- (450.83,106.99) -- (440.13,106.99) -- cycle ;
			\draw  [fill={rgb, 255:red, 173; green, 173; blue, 173 }  ,fill opacity=1 ] (459.43,96.99) -- (470.13,96.99) -- (466.83,107.99) -- (456.13,107.99) -- cycle ;
			\draw  [fill={rgb, 255:red, 173; green, 173; blue, 173 }  ,fill opacity=1 ] (424.3,110) -- (435,110) -- (431.7,121) -- (421,121) -- cycle ;
			\draw  [fill={rgb, 255:red, 173; green, 173; blue, 173 }  ,fill opacity=1 ] (439.43,109.99) -- (450.13,109.99) -- (446.83,120.99) -- (436.13,120.99) -- cycle ;
			\draw  [draw opacity=0][fill={rgb, 255:red, 201; green, 56; blue, 73 }  ,fill opacity=0.72 ] (462.83,63.99) .. controls (471.65,70.64) and (471.71,85.44) .. (462.97,97.04) .. controls (454.23,108.64) and (439.99,112.65) .. (431.17,106.01) .. controls (422.35,99.36) and (422.29,84.56) .. (431.03,72.96) .. controls (439.77,61.36) and (454.01,57.35) .. (462.83,63.99) -- cycle ;
			\draw (218,401.5) node [rotate=-0.96] {\includegraphics[width=102pt,height=80.25pt]{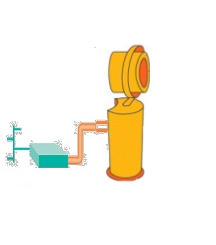}};
			\draw (71,111.5) node  {\includegraphics[width=102pt,height=80.25pt]{8lNWZzncbV-fso.png}};
			\draw  [draw opacity=0][fill={rgb, 255:red, 201; green, 56; blue, 73 }  ,fill opacity=0.72 ] (93.04,89.45) -- (449.41,59.71) -- (450.44,101.74) -- cycle ;
			\draw  [draw opacity=0][fill={rgb, 255:red, 201; green, 56; blue, 73 }  ,fill opacity=0.72 ] (280.09,182.91) -- (440.94,68.35) .. controls (445.18,65.33) and (453.33,69.51) .. (459.16,77.69) .. controls (464.98,85.86) and (466.27,94.94) .. (462.03,97.95) -- (301.18,212.52) .. controls (296.94,215.53) and (288.79,211.35) .. (282.97,203.18) .. controls (277.14,195) and (275.86,185.93) .. (280.09,182.91) .. controls (284.33,179.9) and (292.48,184.08) .. (298.31,192.25) .. controls (304.13,200.43) and (305.42,209.5) .. (301.18,212.52) ;
			\draw  [fill={rgb, 255:red, 198; green, 216; blue, 238 }  ,fill opacity=1 ] (384,350) .. controls (390.63,327.91) and (411.67,310) .. (431,310) .. controls (450.33,310) and (460.63,327.91) .. (454,350) .. controls (447.37,372.09) and (426.33,390) .. (407,390) .. controls (387.67,390) and (377.37,372.09) .. (384,350) -- cycle ;
			\draw  [fill={rgb, 255:red, 173; green, 173; blue, 173 }  ,fill opacity=1 ] (408.3,317) -- (419,317) -- (415.7,328) -- (405,328) -- cycle ;
			\draw  [fill={rgb, 255:red, 173; green, 173; blue, 173 }  ,fill opacity=1 ] (423.43,316.99) -- (434.13,316.99) -- (430.83,327.99) -- (420.13,327.99) -- cycle ;
			\draw  [fill={rgb, 255:red, 173; green, 173; blue, 173 }  ,fill opacity=1 ] (439.43,317.99) -- (450.13,317.99) -- (446.83,328.99) -- (436.13,328.99) -- cycle ;
			\draw  [fill={rgb, 255:red, 173; green, 173; blue, 173 }  ,fill opacity=1 ] (394.43,331.99) -- (405.13,331.99) -- (401.83,342.99) -- (391.13,342.99) -- cycle ;
			\draw  [fill={rgb, 255:red, 173; green, 173; blue, 173 }  ,fill opacity=1 ] (408.43,331.99) -- (419.13,331.99) -- (415.83,342.99) -- (405.13,342.99) -- cycle ;
			\draw  [fill={rgb, 255:red, 173; green, 173; blue, 173 }  ,fill opacity=1 ] (423.43,331.99) -- (434.13,331.99) -- (430.83,342.99) -- (420.13,342.99) -- cycle ;
			\draw  [fill={rgb, 255:red, 173; green, 173; blue, 173 }  ,fill opacity=1 ] (439.43,332.99) -- (450.13,332.99) -- (446.83,343.99) -- (436.13,343.99) -- cycle ;
			\draw  [fill={rgb, 255:red, 173; green, 173; blue, 173 }  ,fill opacity=1 ] (390.43,345.99) -- (401.13,345.99) -- (397.83,356.99) -- (387.13,356.99) -- cycle ;
			\draw  [fill={rgb, 255:red, 173; green, 173; blue, 173 }  ,fill opacity=1 ] (404.43,345.99) -- (415.13,345.99) -- (411.83,356.99) -- (401.13,356.99) -- cycle ;
			\draw  [fill={rgb, 255:red, 173; green, 173; blue, 173 }  ,fill opacity=1 ] (419.43,345.99) -- (430.13,345.99) -- (426.83,356.99) -- (416.13,356.99) -- cycle ;
			\draw  [fill={rgb, 255:red, 173; green, 173; blue, 173 }  ,fill opacity=1 ] (435.43,346.99) -- (446.13,346.99) -- (442.83,357.99) -- (432.13,357.99) -- cycle ;
			\draw  [fill={rgb, 255:red, 173; green, 173; blue, 173 }  ,fill opacity=1 ] (386.43,360.99) -- (397.13,360.99) -- (393.83,371.99) -- (383.13,371.99) -- cycle ;
			\draw  [fill={rgb, 255:red, 173; green, 173; blue, 173 }  ,fill opacity=1 ] (400.43,360.99) -- (411.13,360.99) -- (407.83,371.99) -- (397.13,371.99) -- cycle ;
			\draw  [fill={rgb, 255:red, 173; green, 173; blue, 173 }  ,fill opacity=1 ] (415.43,360.99) -- (426.13,360.99) -- (422.83,371.99) -- (412.13,371.99) -- cycle ;
			\draw  [fill={rgb, 255:red, 173; green, 173; blue, 173 }  ,fill opacity=1 ] (431.43,361.99) -- (442.13,361.99) -- (438.83,372.99) -- (428.13,372.99) -- cycle ;
			\draw  [fill={rgb, 255:red, 173; green, 173; blue, 173 }  ,fill opacity=1 ] (396.3,375) -- (407,375) -- (403.7,386) -- (393,386) -- cycle ;
			\draw  [fill={rgb, 255:red, 173; green, 173; blue, 173 }  ,fill opacity=1 ] (411.43,374.99) -- (422.13,374.99) -- (418.83,385.99) -- (408.13,385.99) -- cycle ;
			\draw  [draw opacity=0][fill={rgb, 255:red, 201; green, 56; blue, 73 }  ,fill opacity=0.72 ] (448.6,310.72) .. controls (465.27,323.27) and (465.52,351.04) .. (449.17,372.74) .. controls (432.83,394.43) and (406.06,401.84) .. (389.4,389.28) .. controls (372.73,376.73) and (372.48,348.96) .. (388.83,327.26) .. controls (405.17,305.57) and (431.94,298.16) .. (448.6,310.72) -- cycle ;
			\draw  [draw opacity=0][fill={rgb, 255:red, 201; green, 56; blue, 73 }  ,fill opacity=0.72 ] (253.46,333.5) -- (413.48,310.02) .. controls (424.89,308.34) and (434.71,324.9) .. (435.43,346.99) .. controls (436.14,369.09) and (427.47,388.36) .. (416.07,390.03) -- (256.04,413.51) .. controls (244.63,415.19) and (234.81,398.63) .. (234.1,376.54) .. controls (233.38,354.44) and (242.05,335.17) .. (253.46,333.5) .. controls (264.86,331.82) and (274.69,348.38) .. (275.4,370.48) .. controls (276.11,392.57) and (267.45,411.84) .. (256.04,413.51) ;
			\draw    (6,283) -- (194,283) ;
			\draw [shift={(196,283)}, rotate = 180] [color={rgb, 255:red, 0; green, 0; blue, 0 }  ][line width=0.75]    (10.93,-3.29) .. controls (6.95,-1.4) and (3.31,-0.3) .. (0,0) .. controls (3.31,0.3) and (6.95,1.4) .. (10.93,3.29)   ;
			\draw    (101,350) -- (101,218) ;
			\draw [shift={(101,216)}, rotate = 450] [color={rgb, 255:red, 0; green, 0; blue, 0 }  ][line width=0.75]    (10.93,-3.29) .. controls (6.95,-1.4) and (3.31,-0.3) .. (0,0) .. controls (3.31,0.3) and (6.95,1.4) .. (10.93,3.29)   ;
			\draw  [fill={rgb, 255:red, 198; green, 216; blue, 238 }  ,fill opacity=0.8 ] (71,283) .. controls (71,266.43) and (84.43,253) .. (101,253) .. controls (117.57,253) and (131,266.43) .. (131,283) .. controls (131,299.57) and (117.57,313) .. (101,313) .. controls (84.43,313) and (71,299.57) .. (71,283) -- cycle ;
			\draw  [fill={rgb, 255:red, 201; green, 56; blue, 73 }  ,fill opacity=0.7 ] (109,250.5) .. controls (109,235.86) and (120.86,224) .. (135.5,224) .. controls (150.14,224) and (162,235.86) .. (162,250.5) .. controls (162,265.14) and (150.14,277) .. (135.5,277) .. controls (120.86,277) and (109,265.14) .. (109,250.5) -- cycle ;
			\draw    (80,264) -- (101,283) ;
			\draw    (135.5,250.5) -- (153,270) ;
			
			\draw (84.77,67.5) node   [align=left] {\LARGE{S}};
			\draw (229.64,442.79) node   [align=left] {\LARGE{D}};
			\draw (457.59,32.5) node   [align=left] {\LARGE{RIS 1}};
			\draw (275.59,147.5) node   [align=left] {\LARGE{RIS 2}};
			\draw (426.51,292.5) node   [align=left] {\LARGE{RIS N-1}};
			\draw (267.02,56.5) node   [align=left] {\LARGE{Link 1}};
			\draw (348.01,121.49) node  [rotate=-324.32] [align=left] {\LARGE{Link 2}};
			\draw (370.9,242.1) node  [font=\LARGE,rotate=-32.55]  {\LARGE{$\cdots $}};
			\draw (46.17,240) node   [align=left] {\begin{minipage}[lt]{52.85pt}\setlength\topsep{0pt}
					\begin{center}
						\LARGE{RIS i-D's}\\\LARGE{plane}
					\end{center}
					
			\end{minipage}};
			\draw (95.56,251.59) node [anchor=north west][inner sep=0.75pt]  [rotate=-41.41]  {\large{$b_{i}$}};
			\draw (150.93,235.12) node [anchor=north west][inner sep=0.75pt]  [rotate=-48.11]  {\large{$w_{d_{i}}$}};

		\end{tikzpicture}

	}
	\vspace{-0.25cm}
	\caption{Cascaded multi-RIS empowered FSO system model.}
	\vspace{-0.8cm}
	\label{Fig:SM}
\end{figure}

As depicted in Fig.~\ref{Fig:SM}, an FSO system, which comprises a source $S$ equipped with a light source (LS), $N-1$ optical RISs, and a destination $D$ equipped with a lens and a photo-detector (PD), is considered. The LS transmits a Gaussian beam towards the RIS. The $i-$th RIS reflects the incident beam towards the ($i+1$)-th RIS until the ($N-1$)-th RIS. The ($N-1$)-th RIS  reflects the incident beam towards the lens $D$, which focuses it to the PD. {Of note, as reported in~\cite{Wang2020}, the RIS is capable of changing the beam shape. We assume that the LS, RISs, and PD are located in fixed positions and that the $i-$th RIS shapes the beam in an optimal manner, i.e. in a way that its footprint at the $i+1-$th RIS and the PD planes are a circle.}   
Finally, it is assumed that the direct $S-D$ link is~blocked, while all the intermediate RIS-empowered and RIS-$D$ links are in line-of-sight~(LoS).

The turbulence coefficient of the $i-$th link is represented by $h_i$, which can be modeled as a GG RV. Each link may or may not experience misalignment fading. Let us assume that $L$ out of $N$ links suffer from misalignment fading and let us model its impact on the $i-$th link through a coefficient $h_{p,i}$. 
 Thus, the received signal at the PD of $D$ can be written~as
\begin{align}
	r = A s + n,\vspace{-0.3cm}
\end{align}       
where $s\in\mathbb{R}^{+}$ represents the intensity of the transmitted signal with $\mathbb{E}[s]\leq P$ with $P$ being the average power constraint, $n\in\mathbf{R}$ stands for the additive white Gaussian shot noise with variance $\sigma_n^2$, which is caused by the ambient light at the PD, and $A\in\mathbb{R}^{+}$ is the end-to-end (e2e) channel. 
The e2e channel coefficient can be expressed~as
\begin{align}
	A = \eta \prod_{i=1}^{L}h_{p,i} h_{i} G_{i} \prod_{i=L+1}^{N} h_{i} G_{i},
	\label{Eq:A}
\end{align}   
where $\eta$ represents the PD responsivity, while $G_i$ stands for the $i$-th link atmospheric gain and can be written~as
\begin{align}
	G_i = \rho 10^{-\alpha \left(d_{i-1}+d_{i}\right)/10}.
	\label{Eq:G}
\end{align} 
In~\eqref{Eq:G}, $\rho$ is the $i-$th RIS reflection efficiency that typically ranges in $[0.7, 1]$~\cite{Huang2016,Ratni2018}, $\alpha$ is a weather-dependent attenuation coefficient,  $d_{i-1}$ and $d_{i}$ are  the distances between the $i-1$ and $i$ links, respectively. Furthermore, the parameters $h_i$ follow the GG distribution  with atmospheric parameters $\alpha_i$ and $\beta_i$,    
calculated based on $d_{i-1}$ or $d_{i+1}$, the refraction index parameter $C_{n}^{2}$ and the wavelength of the optical carrier $\lambda$~\cite{B:Andrews} according to the following formulas 
\begin{align}
	\alpha_i=\left(\exp\left\{ \frac{0.49\sigma_{R_{i}}^{2}}{\left(1+1.11\sigma_{R_{i}}^{\frac{12}{5}}\right)^{\frac{7}{6}}}\right\} -1\right)^{-1}
	\label{Eq:alpha_i}
\end{align}
and
\begin{align}
	\beta_i=\left(\exp\left\{ \frac{0.51\sigma_{R_{i}}^{2}}{\left(1+0.69\sigma_{R_{i}}^{\frac{12}{5}}\right)^{\frac{7}{6}}}\right\} -1\right)^{-1}.
	\label{Eq:beta_i}
\end{align}
Moreover, $\sigma_{R_{i}}^{2}$ is the Rytov variance given by 
\begin{equation}
\sigma_{R_{i}}^{2}=1.23C_{n}^{2}\left(\frac{2\pi}{\lambda}\right)^{\frac{7}{6}}d_{i}^{\frac{11}{6}},
\label{Eq:sigma_R}
\end{equation}
with $C_{n}^2$ being the reflective index structure parameter, which is used to characterize the atmospheric turbulence.   

Finally, the PDF $h_{p,i}$ can be obtained as in~\eqref{Eq:f_l_i}. In this scenario, 
\begin{align}
	A_{o,i} &= [\erf\left(\upsilon_i\right)]^2
	\label{Eq:A_0}
\end{align}
stands for the fraction of the collected power in the ideal case of zero radial displacement. 
Moreover,    
\begin{align}
	\upsilon_i = \frac{\sqrt{\pi}b_i}{\sqrt{2}w_{d_i}}.
	\label{Eq:v}
\end{align}
with $b_i$ and $w_{d_i}$  representing the radius of the circular aperture at the $i-$th RIS, with $i\in[1, N-1]$, or $D$'s side, for $i=N$, respectively and the beam waste  on the corresponding plane. 
Likewise,  $\xi_i$ is the equivalent beam radius, $w_{\mathrm{eq},i}$, to the pointing error displacement standard deviation square ratio and can be calculated~as 
\begin{align}
	\xi_i=\frac{w_{\mathrm{eq},i}^2}{4\sigma_{s,i}^2}
	\label{Eq:xi}
\end{align}
with $\sigma_{s,i}^2$ denoting  the pointing error displacement (jitter) variance. 
Finally,
\begin{align}
	w_{\mathrm{eq},i}^2 &= w_{d_i}^2\frac{\sqrt{\pi}\erf\left(\upsilon_i\right)}{2\upsilon_i\exp\left(-\upsilon_i^2\right)}.
	\label{Eq:wZeq}
\end{align}

\subsubsection{Performance assessment}

The SNR at the RX can be expressed~as
\begin{align}
	\rho_{\text{FSO}} = A^2 \rho_s,
	\label{Eq:rho_FSO}
\end{align}
where $\rho_s$ stands for the transmission SNR multiplied by the deterministic path-gain. Thus, the OP is defined~as
\begin{align}
P_o^{\text{FSO}} = \Pr\left(\rho_{\text{FSO}} \leq r_{\text{th}}\right), 	
\end{align}
or, from~\eqref{Eq:rho_FSO}, 
\begin{align}
	P_o^{\text{FSO}} = \Pr\left(A \leq \sqrt{\frac{r_{\text{th}}}{\rho_s}}\right),
	\label{Eq:P_o_FSO} 	
\end{align}
which, by applying~\eqref{Eq:F_Z_final}, can be written in a closed-form expression. {For the special case, in which $d_1\neq d_2\neq\cdots\neq d_N$,~\eqref{Eq:FZalternative} can be employed in order to asses the OP of the cascaded multi-RIS empowered FSO system.} 
Of note, for the special case in which $N=L=2$, $d_1=d_2$, $b_1=b_2$, $w_{d_1}=w_{d_2}$ and $\sigma_{s,1}=\sigma_{s,2}$, \eqref{Eq:P_o_FSO} can be written as in~\cite[eq. (15)]{Ndjiongue2021}.

{The following lemma returns the diversity order of the cascaded multi-RIS empowered FSO system.
\begin{lem}
	The diversity order of the cascaded multi-RIS empowered FSO system can be obtained~as
	\begin{align}
		D_{\text{FSO}} = \min_{i=1,\cdots,2N+L}\mathcal{B}_i/2.
		\label{Eq:D}
	\end{align}
\end{lem}
\begin{IEEEproof}
	For~$\rho_s\to\infty$, the term $\frac{r_{\mathrm{th}}}{\rho_s}$ tend to $0$; thus, based on~\eqref{Eq:FZalternative_to_0}, the outage probability can be approximated~as
	\begin{align}
		&P_{o,\infty}^{\text{FSO}} = \frac{\prod_{i=1}^{L}\xi_i}{\prod_{i=1}^N\Gamma\left(\alpha_i\right) \Gamma\left(\beta_i\right)} 
		\nonumber \\ & \times
		\sum_{i=1}^{2N+L} \frac{\prod_{j=1}^{2N+L}\Gamma\left(\mathcal{B}_j-\mathcal{B}_i\right)\Gamma\left(\mathcal{A}_i\right)}{\prod_{j=2N+L}^{2N+L+1}
			\Gamma\left(1+\mathcal{B}_i-\mathcal{B}_j\right)\prod_{j=2}^{L+1}\Gamma\left(\mathcal{A}_j-\mathcal{B}_i\right)}
			\nonumber \\ & \hspace{+5.5cm}\times
		 \left(\frac{r_{\mathrm{th}}}{\rho_{s}}\right)^{\mathcal{B}_i/2}.
		 \label{Eq:P_o_infty_FSO}
	\end{align}
It is evident that in~\eqref{Eq:P_o_infty_FSO}, $\left(\frac{r_{\mathrm{th}}}{\rho_{s}}\right)^{\mathcal{B}_i/2}$ contributes with diversity order $\mathcal{B}_i/2$ in the asymptotic OP. Hence, the diversity order can be obtained as in~\eqref{Eq:D}. This concludes the proof. 
\end{IEEEproof}
}

\vspace{-0.3cm}\subsection{Parallel Multi-RIS FSO systems} \label{SS:FSO2} \vspace{-0.2cm}

\subsubsection{System model}

\begin{figure}
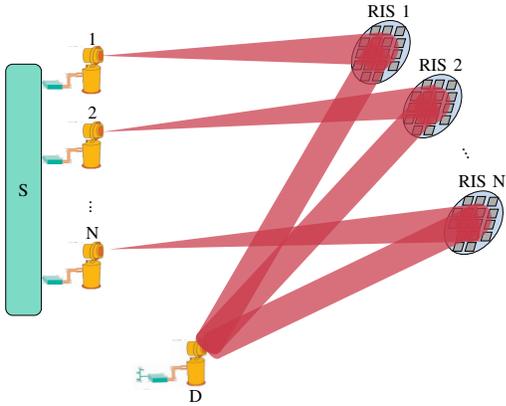

	\centering
	\scalebox{0.4}{
		
		\tikzset{every picture/.style={line width=0.75pt}} 
		
		\begin{tikzpicture}[x=0.75pt,y=0.75pt,yscale=-1,xscale=1]
			
			\draw (257,501.5) node [rotate=-358.54] {\includegraphics[width=102pt,height=80.25pt]{8lNWZzncbV-fso.png}};
			\draw  [fill={rgb, 255:red, 198; green, 216; blue, 238 }  ,fill opacity=1 ] (466,105) .. controls (472.63,82.91) and (493.67,65) .. (513,65) .. controls (532.33,65) and (542.63,82.91) .. (536,105) .. controls (529.37,127.09) and (508.33,145) .. (489,145) .. controls (469.67,145) and (459.37,127.09) .. (466,105) -- cycle ;
			\draw  [fill={rgb, 255:red, 173; green, 173; blue, 173 }  ,fill opacity=1 ] (490.3,72) -- (501,72) -- (497.7,83) -- (487,83) -- cycle ;
			\draw  [fill={rgb, 255:red, 173; green, 173; blue, 173 }  ,fill opacity=1 ] (505.43,71.99) -- (516.13,71.99) -- (512.83,82.99) -- (502.13,82.99) -- cycle ;
			\draw  [fill={rgb, 255:red, 173; green, 173; blue, 173 }  ,fill opacity=1 ] (521.43,72.99) -- (532.13,72.99) -- (528.83,83.99) -- (518.13,83.99) -- cycle ;
			\draw  [fill={rgb, 255:red, 173; green, 173; blue, 173 }  ,fill opacity=1 ] (476.43,86.99) -- (487.13,86.99) -- (483.83,97.99) -- (473.13,97.99) -- cycle ;
			\draw  [fill={rgb, 255:red, 173; green, 173; blue, 173 }  ,fill opacity=1 ] (490.43,86.99) -- (501.13,86.99) -- (497.83,97.99) -- (487.13,97.99) -- cycle ;
			\draw  [fill={rgb, 255:red, 173; green, 173; blue, 173 }  ,fill opacity=1 ] (505.43,86.99) -- (516.13,86.99) -- (512.83,97.99) -- (502.13,97.99) -- cycle ;
			\draw  [fill={rgb, 255:red, 173; green, 173; blue, 173 }  ,fill opacity=1 ] (521.43,87.99) -- (532.13,87.99) -- (528.83,98.99) -- (518.13,98.99) -- cycle ;
			\draw  [fill={rgb, 255:red, 173; green, 173; blue, 173 }  ,fill opacity=1 ] (472.43,100.99) -- (483.13,100.99) -- (479.83,111.99) -- (469.13,111.99) -- cycle ;
			\draw  [fill={rgb, 255:red, 173; green, 173; blue, 173 }  ,fill opacity=1 ] (486.43,100.99) -- (497.13,100.99) -- (493.83,111.99) -- (483.13,111.99) -- cycle ;
			\draw  [fill={rgb, 255:red, 173; green, 173; blue, 173 }  ,fill opacity=1 ] (501.43,100.99) -- (512.13,100.99) -- (508.83,111.99) -- (498.13,111.99) -- cycle ;
			\draw  [fill={rgb, 255:red, 173; green, 173; blue, 173 }  ,fill opacity=1 ] (517.43,101.99) -- (528.13,101.99) -- (524.83,112.99) -- (514.13,112.99) -- cycle ;
			\draw  [fill={rgb, 255:red, 173; green, 173; blue, 173 }  ,fill opacity=1 ] (468.43,115.99) -- (479.13,115.99) -- (475.83,126.99) -- (465.13,126.99) -- cycle ;
			\draw  [fill={rgb, 255:red, 173; green, 173; blue, 173 }  ,fill opacity=1 ] (482.43,115.99) -- (493.13,115.99) -- (489.83,126.99) -- (479.13,126.99) -- cycle ;
			\draw  [fill={rgb, 255:red, 173; green, 173; blue, 173 }  ,fill opacity=1 ] (497.43,115.99) -- (508.13,115.99) -- (504.83,126.99) -- (494.13,126.99) -- cycle ;
			\draw  [fill={rgb, 255:red, 173; green, 173; blue, 173 }  ,fill opacity=1 ] (513.43,116.99) -- (524.13,116.99) -- (520.83,127.99) -- (510.13,127.99) -- cycle ;
			\draw  [fill={rgb, 255:red, 173; green, 173; blue, 173 }  ,fill opacity=1 ] (478.3,130) -- (489,130) -- (485.7,141) -- (475,141) -- cycle ;
			\draw  [fill={rgb, 255:red, 173; green, 173; blue, 173 }  ,fill opacity=1 ] (493.43,129.99) -- (504.13,129.99) -- (500.83,140.99) -- (490.13,140.99) -- cycle ;
			\draw  [draw opacity=0][fill={rgb, 255:red, 201; green, 56; blue, 73 }  ,fill opacity=0.72 ] (516.83,83.99) .. controls (525.65,90.64) and (525.71,105.44) .. (516.97,117.04) .. controls (508.23,128.64) and (493.99,132.65) .. (485.17,126.01) .. controls (476.35,119.36) and (476.29,104.56) .. (485.03,92.96) .. controls (493.77,81.36) and (508.01,77.35) .. (516.83,83.99) -- cycle ;
			\draw (125,131.5) node  {\includegraphics[width=102pt,height=80.25pt]{8lNWZzncbV-fso.png}};
			\draw  [draw opacity=0][fill={rgb, 255:red, 201; green, 56; blue, 73 }  ,fill opacity=0.72 ] (147.04,109.45) -- (503.41,79.71) -- (504.44,121.74) -- cycle ;
			\draw  [draw opacity=0][fill={rgb, 255:red, 201; green, 56; blue, 73 }  ,fill opacity=0.72 ] (268.21,463.48) -- (481.69,96.98) .. controls (484.3,92.49) and (493.46,92.94) .. (502.13,97.99) .. controls (510.8,103.05) and (515.71,110.78) .. (513.09,115.28) -- (299.62,481.77) .. controls (297,486.26) and (287.85,485.81) .. (279.17,480.76) .. controls (270.5,475.71) and (265.59,467.97) .. (268.21,463.48) .. controls (270.83,458.98) and (279.98,459.44) .. (288.65,464.49) .. controls (297.33,469.54) and (302.23,477.28) .. (299.62,481.77) ;
			\draw (125,226.5) node  {\includegraphics[width=102pt,height=80.25pt]{8lNWZzncbV-fso.png}};
			\draw  [fill={rgb, 255:red, 198; green, 216; blue, 238 }  ,fill opacity=1 ] (531,173) .. controls (537.63,150.91) and (558.67,133) .. (578,133) .. controls (597.33,133) and (607.63,150.91) .. (601,173) .. controls (594.37,195.09) and (573.33,213) .. (554,213) .. controls (534.67,213) and (524.37,195.09) .. (531,173) -- cycle ;
			\draw  [fill={rgb, 255:red, 173; green, 173; blue, 173 }  ,fill opacity=1 ] (555.3,140) -- (566,140) -- (562.7,151) -- (552,151) -- cycle ;
			\draw  [fill={rgb, 255:red, 173; green, 173; blue, 173 }  ,fill opacity=1 ] (570.43,139.99) -- (581.13,139.99) -- (577.83,150.99) -- (567.13,150.99) -- cycle ;
			\draw  [fill={rgb, 255:red, 173; green, 173; blue, 173 }  ,fill opacity=1 ] (586.43,140.99) -- (597.13,140.99) -- (593.83,151.99) -- (583.13,151.99) -- cycle ;
			\draw  [fill={rgb, 255:red, 173; green, 173; blue, 173 }  ,fill opacity=1 ] (541.43,154.99) -- (552.13,154.99) -- (548.83,165.99) -- (538.13,165.99) -- cycle ;
			\draw  [fill={rgb, 255:red, 173; green, 173; blue, 173 }  ,fill opacity=1 ] (555.43,154.99) -- (566.13,154.99) -- (562.83,165.99) -- (552.13,165.99) -- cycle ;
			\draw  [fill={rgb, 255:red, 173; green, 173; blue, 173 }  ,fill opacity=1 ] (570.43,154.99) -- (581.13,154.99) -- (577.83,165.99) -- (567.13,165.99) -- cycle ;
			\draw  [fill={rgb, 255:red, 173; green, 173; blue, 173 }  ,fill opacity=1 ] (586.43,155.99) -- (597.13,155.99) -- (593.83,166.99) -- (583.13,166.99) -- cycle ;
			\draw  [fill={rgb, 255:red, 173; green, 173; blue, 173 }  ,fill opacity=1 ] (537.43,168.99) -- (548.13,168.99) -- (544.83,179.99) -- (534.13,179.99) -- cycle ;
			\draw  [fill={rgb, 255:red, 173; green, 173; blue, 173 }  ,fill opacity=1 ] (551.43,168.99) -- (562.13,168.99) -- (558.83,179.99) -- (548.13,179.99) -- cycle ;
			\draw  [fill={rgb, 255:red, 173; green, 173; blue, 173 }  ,fill opacity=1 ] (566.43,168.99) -- (577.13,168.99) -- (573.83,179.99) -- (563.13,179.99) -- cycle ;
			\draw  [fill={rgb, 255:red, 173; green, 173; blue, 173 }  ,fill opacity=1 ] (582.43,169.99) -- (593.13,169.99) -- (589.83,180.99) -- (579.13,180.99) -- cycle ;
			\draw  [fill={rgb, 255:red, 173; green, 173; blue, 173 }  ,fill opacity=1 ] (533.43,183.99) -- (544.13,183.99) -- (540.83,194.99) -- (530.13,194.99) -- cycle ;
			\draw  [fill={rgb, 255:red, 173; green, 173; blue, 173 }  ,fill opacity=1 ] (547.43,183.99) -- (558.13,183.99) -- (554.83,194.99) -- (544.13,194.99) -- cycle ;
			\draw  [fill={rgb, 255:red, 173; green, 173; blue, 173 }  ,fill opacity=1 ] (562.43,183.99) -- (573.13,183.99) -- (569.83,194.99) -- (559.13,194.99) -- cycle ;
			\draw  [fill={rgb, 255:red, 173; green, 173; blue, 173 }  ,fill opacity=1 ] (578.43,184.99) -- (589.13,184.99) -- (585.83,195.99) -- (575.13,195.99) -- cycle ;
			\draw  [fill={rgb, 255:red, 173; green, 173; blue, 173 }  ,fill opacity=1 ] (543.3,198) -- (554,198) -- (550.7,209) -- (540,209) -- cycle ;
			\draw  [fill={rgb, 255:red, 173; green, 173; blue, 173 }  ,fill opacity=1 ] (558.43,197.99) -- (569.13,197.99) -- (565.83,208.99) -- (555.13,208.99) -- cycle ;
			\draw  [draw opacity=0][fill={rgb, 255:red, 201; green, 56; blue, 73 }  ,fill opacity=0.72 ] (581.83,151.99) .. controls (590.65,158.64) and (590.71,173.44) .. (581.97,185.04) .. controls (573.23,196.64) and (558.99,200.65) .. (550.17,194.01) .. controls (541.35,187.36) and (541.29,172.56) .. (550.03,160.96) .. controls (558.77,149.36) and (573.01,145.35) .. (581.83,151.99) -- cycle ;
			\draw  [draw opacity=0][fill={rgb, 255:red, 201; green, 56; blue, 73 }  ,fill opacity=0.72 ] (149.1,205.54) -- (570.06,145.58) -- (573.91,187.18) -- cycle ;
			\draw  [draw opacity=0][fill={rgb, 255:red, 201; green, 56; blue, 73 }  ,fill opacity=0.72 ] (269.62,459.88) -- (547.4,160.53) .. controls (550.94,156.72) and (559.77,159.17) .. (567.13,165.99) .. controls (574.49,172.82) and (577.58,181.45) .. (574.05,185.26) -- (296.26,484.61) .. controls (292.73,488.42) and (283.89,485.97) .. (276.54,479.15) .. controls (269.18,472.32) and (266.08,463.69) .. (269.62,459.88) .. controls (273.16,456.07) and (281.99,458.52) .. (289.35,465.34) .. controls (296.7,472.17) and (299.8,480.8) .. (296.26,484.61) ;
			\draw  [fill={rgb, 255:red, 198; green, 216; blue, 238 }  ,fill opacity=1 ] (583,320) .. controls (589.63,297.91) and (610.67,280) .. (630,280) .. controls (649.33,280) and (659.63,297.91) .. (653,320) .. controls (646.37,342.09) and (625.33,360) .. (606,360) .. controls (586.67,360) and (576.37,342.09) .. (583,320) -- cycle ;
			\draw  [fill={rgb, 255:red, 173; green, 173; blue, 173 }  ,fill opacity=1 ] (607.3,287) -- (618,287) -- (614.7,298) -- (604,298) -- cycle ;
			\draw  [fill={rgb, 255:red, 173; green, 173; blue, 173 }  ,fill opacity=1 ] (622.43,286.99) -- (633.13,286.99) -- (629.83,297.99) -- (619.13,297.99) -- cycle ;
			\draw  [fill={rgb, 255:red, 173; green, 173; blue, 173 }  ,fill opacity=1 ] (638.43,287.99) -- (649.13,287.99) -- (645.83,298.99) -- (635.13,298.99) -- cycle ;
			\draw  [fill={rgb, 255:red, 173; green, 173; blue, 173 }  ,fill opacity=1 ] (593.43,301.99) -- (604.13,301.99) -- (600.83,312.99) -- (590.13,312.99) -- cycle ;
			\draw  [fill={rgb, 255:red, 173; green, 173; blue, 173 }  ,fill opacity=1 ] (607.43,301.99) -- (618.13,301.99) -- (614.83,312.99) -- (604.13,312.99) -- cycle ;
			\draw  [fill={rgb, 255:red, 173; green, 173; blue, 173 }  ,fill opacity=1 ] (622.43,301.99) -- (633.13,301.99) -- (629.83,312.99) -- (619.13,312.99) -- cycle ;
			\draw  [fill={rgb, 255:red, 173; green, 173; blue, 173 }  ,fill opacity=1 ] (638.43,302.99) -- (649.13,302.99) -- (645.83,313.99) -- (635.13,313.99) -- cycle ;
			\draw  [fill={rgb, 255:red, 173; green, 173; blue, 173 }  ,fill opacity=1 ] (589.43,315.99) -- (600.13,315.99) -- (596.83,326.99) -- (586.13,326.99) -- cycle ;
			\draw  [fill={rgb, 255:red, 173; green, 173; blue, 173 }  ,fill opacity=1 ] (603.43,315.99) -- (614.13,315.99) -- (610.83,326.99) -- (600.13,326.99) -- cycle ;
			\draw  [fill={rgb, 255:red, 173; green, 173; blue, 173 }  ,fill opacity=1 ] (618.43,315.99) -- (629.13,315.99) -- (625.83,326.99) -- (615.13,326.99) -- cycle ;
			\draw  [fill={rgb, 255:red, 173; green, 173; blue, 173 }  ,fill opacity=1 ] (634.43,316.99) -- (645.13,316.99) -- (641.83,327.99) -- (631.13,327.99) -- cycle ;
			\draw  [fill={rgb, 255:red, 173; green, 173; blue, 173 }  ,fill opacity=1 ] (599.43,330.99) -- (610.13,330.99) -- (606.83,341.99) -- (596.13,341.99) -- cycle ;
			\draw  [fill={rgb, 255:red, 173; green, 173; blue, 173 }  ,fill opacity=1 ] (614.43,330.99) -- (625.13,330.99) -- (621.83,341.99) -- (611.13,341.99) -- cycle ;
			\draw  [fill={rgb, 255:red, 173; green, 173; blue, 173 }  ,fill opacity=1 ] (630.43,331.99) -- (641.13,331.99) -- (637.83,342.99) -- (627.13,342.99) -- cycle ;
			\draw  [fill={rgb, 255:red, 173; green, 173; blue, 173 }  ,fill opacity=1 ] (595.3,345) -- (606,345) -- (602.7,356) -- (592,356) -- cycle ;
			\draw  [fill={rgb, 255:red, 173; green, 173; blue, 173 }  ,fill opacity=1 ] (610.43,344.99) -- (621.13,344.99) -- (617.83,355.99) -- (607.13,355.99) -- cycle ;
			\draw  [draw opacity=0][fill={rgb, 255:red, 201; green, 56; blue, 73 }  ,fill opacity=0.72 ] (633.83,298.99) .. controls (642.65,305.64) and (642.71,320.44) .. (633.97,332.04) .. controls (625.23,343.64) and (610.99,347.65) .. (602.17,341.01) .. controls (593.35,334.36) and (593.29,319.56) .. (602.03,307.96) .. controls (610.77,296.36) and (625.01,292.35) .. (633.83,298.99) -- cycle ;
			\draw  [draw opacity=0][fill={rgb, 255:red, 201; green, 56; blue, 73 }  ,fill opacity=0.72 ] (158.35,353.71) -- (622.77,296.37) -- (626.37,343.76) -- cycle ;
			\draw  [draw opacity=0][fill={rgb, 255:red, 201; green, 56; blue, 73 }  ,fill opacity=0.72 ] (281.01,458.6) -- (607.43,301.99) .. controls (612.12,299.75) and (619.44,305.26) .. (623.78,314.31) .. controls (628.12,323.36) and (627.84,332.52) .. (623.15,334.77) -- (296.73,491.37) .. controls (292.05,493.62) and (284.73,488.1) .. (280.38,479.06) .. controls (276.04,470.01) and (276.32,460.85) .. (281.01,458.6) .. controls (285.7,456.35) and (293.02,461.86) .. (297.36,470.91) .. controls (301.7,479.96) and (301.42,489.12) .. (296.73,491.37) ;
			\draw (125,378.5) node  {\includegraphics[width=102pt,height=80.25pt]{8lNWZzncbV-fso.png}};
			\draw  [fill={rgb, 255:red, 124; green, 218; blue, 197 }  ,fill opacity=1 ] (28,129) .. controls (28,124.03) and (32.03,120) .. (37,120) -- (64,120) .. controls (68.97,120) and (73,124.03) .. (73,129) -- (73,428) .. controls (73,432.97) and (68.97,437) .. (64,437) -- (37,437) .. controls (32.03,437) and (28,432.97) .. (28,428) -- cycle ;
			
			\draw (50.5,278.5) node   [align=left] {\LARGE{S}};
			\draw (511.59,52.5) node   [align=left] {\LARGE{RIS 1}};
			\draw (576.59,120.5) node   [align=left] {\LARGE{RIS 2}};
			\draw (628.59,267.5) node   [align=left] {\LARGE{RIS N}};
			\draw (267.5,538.5) node   [align=left] {\LARGE{D}};
			\draw (606.23,229.12) node  [font=\Large,rotate=-329.14]  {\LARGE{$\vdots $}};
			\draw (135.21,296.07) node  [font=\Large]  {\LARGE{$\vdots $}};
			\draw (137.5,87.5) node   [align=left] {\LARGE{1}};
			\draw (137.5,180.5) node   [align=left] {\LARGE{2}};
			\draw (137.5,332.5) node   [align=left] {\LARGE{N}};

		\end{tikzpicture}
	}
	\vspace{-0.25cm}
	\caption{Parallel Multi-RIS FSO system model.}
	\vspace{-0.8cm}
	\label{Fig:SM2}
\end{figure}

As presented in Fig.~\ref{Fig:SM2}, we consider a multi-RIS FSO system that consists of a $S$, $N$ RIS, and a $D$. The $S$ is equipped with $N$ transmit apertures, each one at a different RIS. The $i-$th RIS steers the incident beam towards $D$, which is equipped with a single PD. Thus, the received signal at the destination can be expressed~as 
\begin{align}
	r_p = \sum_{i=1}^{N} B_{i} s_i +n,
\end{align}
where $s_i$ is the transmission signal by the $i-$th S's aperture,~while
 \begin{align}
 	B_i = h_{p_{1,i}} h_{p_{2,i}} h_{1,i} h_{2,i} G_{i},
 	\label{Eq:Bi}
 \end{align}
represents the $S$-$i-$th RIS-$D$ channel coefficient. In more detail, in~\eqref{Eq:Bi}, $h_{p_{1,i}}$ and~$h_{p_{2,i}}$  denote the fading misalignment coefficient of the $S$-$i-$th RIS and $i-$th RIS-$D$ links, respectively, whereas $h_{1,i}$ and $h_{2,i}$ are the turbulence coefficients of the $S$-$i-$th RIS and $i-$th RIS-$D$~links, respectively.

\subsubsection{Performance assessment}

By assuming that the total transmission power, $P_s$, is equally divided into the $N$ transmission apertures, the instantaneous SNR at $D$ can be obtained~as
\begin{align}
	\rho_{\text{par}} = \frac{\left(\sum_{i=1}^{N} B_{i}\right)^2 P_s}{N N_o},
\end{align}   
or equivalent
\begin{align}
	\rho_{\text{par}} = N S^2 \rho_s,
	\label{Eq:rho_par}
\end{align}
where 
\begin{align}
	S = \frac{1}{N}\sum_{i=1}^{N} B_{i}. 
\end{align}

From~\eqref{Eq:rho_par}, the OP can be expressed~as
\begin{align}
	P_o^{\text{par}} = \Pr\left(\rho_{\text{par}}\leq \rho_{\mathrm{th}}\right),
\end{align}
or equivalently
\begin{align}
	P_o^{\text{par}} = \Pr\left(S\leq \sqrt{\frac{\rho_{\mathrm{th}}}{N\rho_s}}\right),
\end{align}
or
\begin{align}
	P_o^{\text{par}} = F_S\left( \sqrt{\frac{\rho_{\mathrm{th}}}{N\rho_s}}\right),
\end{align}
with $F_S\left(\cdot\right)$ being the CDF of $S$. According to~\cite[eq.(14)]{Karagiannidis2006}, the OP of the parallel multi-RIS FSO scenario can be upper bounded~as  
\begin{align}
	P_o^{\text{par}} \leq F_A\left( \left(\sqrt{\frac{\rho_{\mathrm{th}}}{N\rho_s}}\right)^{N}\right).
	\label{Eq:Po_bound}
\end{align}

\vspace{-0.3cm}
\subsection{RIS-empowered THz wireless systems}
\vspace{-0.1cm}
{In this section, the system model and theoretical framework that quantifies the outage performance of multi-RIS-empowered outdoor THz wireless systems is provided. Note that this analysis can be also and straightforwardly extended to millimeter wave systems by replacing the term that describes the deterministic path-gain of the THz system with the corresponding one for the millimeter wave system.}
\subsubsection{System model}
A multi-RIS-empowered outdoor THz wireless systems is considered, in which $S$ and $D$ are equipped with high-directional antennas. {This system can be used a wireless fiber extender; thus, both the $S$ and $D$ antennas are placed high, e.g., in rooftops.}
Again, it is assumed that {due to static obstacles, e.g., buildings}\footnote{{Note that, since the transmission and reception antennas as well as the RISs are placed high, no dynamic blockage is expected.}}, no direct link between the $S$ and $D$ can be established; as a result, $N$ RISs are used. Each RIS can be seen as a reflector that steers and/or focuses the beams towards the desired direction. Thus, the received signal at $D$ can be obtained~as
\begin{align}
	y = A_{t} \left(x + \eta_{s}\right) + \eta_{d} + w,
	\label{Eq:y}
\end{align}    
where $x$ is the transmission signal, while $w$ stands for the AWGN. Moreover, $\eta_{s}$ and $\eta_{d}$ are the $S$ and $D$ distortion noises, due to transmitter's and receiver's hardware imperfections, respectively. According to~\cite{Boulogeorgos2020b,ref14_Al_hard_imperf,B:Schenk-book,Boulogeorgos2018a}, $\eta_{s}$ and $\eta_{d}$ can be modeled as two independent RVs that, for a given channel realization, have zero-mean complex Gaussian distributions with variances
\begin{align} 
	\sigma_{s}^2=\kappa_{s}^2 P_s \text{ and } \sigma_{d}^2=\kappa_{r}^2 A^2 P_s,
	\label{Eq:sigma_of_distortion}
\end{align} 
respectively. In~\eqref{Eq:sigma_of_distortion}, $\kappa_{t}$ and $\kappa_{r}$ stand for the error vector magnitude of the S's transmitter and the $D$'s receiver, while $P_s$ denotes the transmission power. 

Meanwhile, in~\eqref{Eq:y}, $A_{t}$ represents the e2e channel, which can be expressed~as  
\begin{align}
	A_{t} = \prod_{i=1}^{L} h_{p,i}^{\text{THz}} h_{t,i}^{\text{THz}} g_i 
	\prod_{i=L+1}^{N} h_{t,i}^{\text{THz}} g_i. 
	\label{Eq:A_t}
\end{align}  
In~\eqref{Eq:A_t}, $h_{p,i}^{\text{THz}}$ and $h_{t,i}^{\text{THz}}$ are  the misalignment fading and turbulence coefficients that are distributed according to~\eqref{Eq:f_l_i} and~\eqref{Eq:f_r_i}, respectively. {Note that the turbulence model that is employed has been experimentally validated in~\cite{Ma2014}, theoretically analyzed in~\cite{Taherkhani2020}, and supported by ITU-R~\cite{ITUR2013}.} The $\alpha_i$ and $\beta_i$ parameters of~\eqref{Eq:f_r_i} can be evaluated according to~\cite{Taherkhani2020}~as
\begin{align}
	\alpha_i^{\text{THz}}\hspace{-0.2cm} &=\hspace{-0.15cm} {\left(\hspace{-0.1cm}\exp\left(\frac{0.49\sigma_{R_i}^2}{\left(1+0.65 D_i^2+1.11\sigma_{R_i}^{12/5}\right)^{7/6}} \right)-1 \hspace{-0.1cm}\right)^{-1}} 
\end{align} 
and
\begin{align}
	\beta_i^{\text{THz}} \hspace{-0.2cm} &=\hspace{-0.15cm} \left(\hspace{-0.1cm}\exp\left( \frac{0.51\sigma_{R_i}^2\left(1+0.69\sigma_{R_i}^{12/5}\right)^{-5/6}}{1+0.9 D_i^2 +0.62 D_i^2 \sigma_{R_i}^{12/5}} \right)-1 \hspace{-0.1cm}\right)^{-1}
\end{align}
while~$\sigma_{R_i}^2$ can be obtained as in~\eqref{Eq:sigma_R}. Additionally, 
\begin{align}
	D_i = \sqrt{\frac{\pi b_i^2}{2\lambda d_i}}.
\end{align}
 According to ITU-T, in the THz band, the reflection index structure parameter  can be formulated as~\cite{Hill1988}
\begin{align}
\hspace{-0.2cm}	C_{n}^{2}=  \frac{C_{T}^2}{T} \left(A_{T}^2\left(\lambda\right) + 10^{4} A_{Q}^{2}\left(\lambda\right) \pm 200 A_{T}\left(\lambda\right) A_{Q}\left(\lambda\right)\right),
\end{align}          
where $T$ stands for the temperature, $C_{T}$ denotes the structure factor for the temperature, while $A_{T}$ and $A_{R}$ are the derivatives of the
real part of the refractive index with respect to the temperature and humidity, respectively. Finally, $g_i$ represents the deterministic path-gain coefficient of the $i-$th link and can be written~as
\begin{align}
	g_i =  g_{f,i}\left(f, d_i\right) \tau\left(f, d_i\right),
\end{align} 
where $g_{f,i}$ denotes the free space path loss coefficient, which based on the Friis transmission equation, can be expressed~as 
\begin{align}
	g_{f,i}\left(f, d_i\right)=\left\{ \begin{array}{l l}
		\frac{c}{4\pi f d_i} \sqrt{G_{s}}, & \text{ for } i=1 \\
		\frac{c}{4\pi f d_i} R_{i-1}, & \text{ for } i\in[2, N-1] \\
		\frac{c}{4\pi f d_i} \sqrt{G_{d}}, & \text{ for } i=N 
	\end{array}\right.,
\label{Eq:g_f}
\end{align}
whereas $\tau\left(f, d_i\right)$ denotes the molecular absorption coefficient and can be obtained~as
\begin{align}
	\tau\left(f, d_i\right) = \exp\left(-\frac{1}{2}\kappa(f)\, d_i\right).
	\label{Eq:tau}
\end{align}
In~\eqref{Eq:g_f}, $f$ and $c$  stand for the transmission frequency and the speed of light, respectively, while  $G_{s}$, $R_{i}$ and $G_{d}$ represent the $S$ transmission antenna gain, the $i-$th RIS reflection coefficient, and the $D$ reception antenna~gain, respectively. Likewise, in~\eqref{Eq:tau}, $\kappa(f)$ denotes the molecular absorption coefficient, which depends on the atmospheric temperature ($T$), pressure ($p$), as well as relative humidity ($\phi$) and {can be evaluated as in~\cite{Kokkoniemi2021,Kokkoniemi2021a} and~\cite[eq. (8)]{A:Analytical_Performance_Assessment_of_THz_Wireless_Systems}
	\begin{align}
		\kappa(f) = g(f) + g_1 + g_2,
	\end{align}
where 
\begin{align}
	g(f) = \sum_{i=0}^{3}p_i f^i,
\end{align}
with $p_0=-6.36$, $p_1=9.06$, $p_2=-3.94$ and $p_3=5.54$. Moreover,
\begin{align}
	g_1 = \frac{G_{A}\left(\mu_w\right)}{G_{B}\left(\mu_w\right) + \left(\frac{f}{100\,c}-c_0\right)^2}, 
\end{align}
where $G_{A}\left(\mu_w\right) = 0.2205\,\mu_w \left(0.1303\mu_w+0.0294\right)$, $G_B\left(\mu_w\right)=\left(0.4093\,\mu_w+0.0925\right)^2$, $c_0=10.835$ with $\mu_w$ being the volume mixing ratio of the water vapor and can be computed~as 
\begin{align}
	\mu_{w}=\frac{\phi}{100}\frac{p_w\left(T, p\right)}{p}.
	\label{Eq:mu_w}
\end{align}
In~\eqref{Eq:mu_w}, $p_w\left(T, p\right)$ is the saturated water vapor partial pressure and can be expressed~as
\begin{align}
	p_w\left(T, p\right) = k_1 \left(k_2 + k_3 \phi_p\right) \exp\left(\frac{k_4\left(T-T_1\right)}{T-T_2}\right),
\end{align}
where $k_1=6.1121$, $k_2=1.0007$, $k_3=3.46\times 10^{-6}\,\mathrm{hPa}^{-1}$, $k_4=17.502$, $T_1=273.15^o\mathrm{K}$, and $T_2=32.18^o\mathrm{K}$. Finally, 
\begin{align}
	g_2 = \frac{G_{C}\left(\mu_w\right)}{G_D\left(\mu_w\right) + \left(\frac{f}{100\,c}-c_1\right)^2},
\end{align} 
where $G_{C}\left(\mu_w\right) = 2.014\,\mu_w \left(0.1702\mu_w+0.0303\right)$, $G_{D}=\left(0.537\, \mu_w + 0.0956\right)^2$ and $c_1=12.664$. 
}

\subsubsection{Performance assessment}
From~\eqref{Eq:y}, the signal-to-distortion-plus-noise-ratio (SDNR) can be  obtained~as 
\begin{align}
	\gamma = \frac{A_{t}^2 P_s}{A_{t}^2 \left(\kappa_t^2 + \kappa_r^2 \right) P_s + N_o},
	\label{Eq:gamma}
\end{align}
where $N_o$ is the noise variance. Moreover, the OP is defined~as
\begin{align}
	P_o^{\text{THz}} = \Pr\left(\gamma\leq \gamma_{\mathrm{th}}\right),
	\label{Eq:P_o_THz}
\end{align} 
where $\gamma_{\mathrm{th}}$ represents the SNR threshold. With the aid of~\eqref{Eq:gamma},~\eqref{Eq:P_o_THz} can be rewritten~as
\begin{align}
	P_o^{\text{THz}} = \left\{\begin{array}{l l} \Pr\left( A_{t}^2 \leq \frac{1}{1-\gamma_{\mathrm{th}}\left(\kappa_t^2 + \kappa_r^2\right)} \frac{\gamma_{\mathrm{th}}}{\gamma_{s}} \right),  & \text{ for } \gamma_{\mathrm{th}} \leq \frac{1}{\kappa_{t}^2 + \kappa_r^2} \\ 1, & \text{ for } \gamma_{\mathrm{th}} > \frac{1}{\kappa_{t}^2 + \kappa_r^2} \end{array}\right..
	\label{Eq:P_o_THz_s2} 
\end{align}
Finally, by applying~\eqref{Eq:A_t} in~\eqref{Eq:P_o_THz_s2}, we get~\eqref{Eq:P_o_THz_s3}, given at the top of the next page.
\begin{figure*}
 \begin{align}
 	P_o^{\text{THz}} = \left\{\begin{array}{l l} F_Z\left( \sqrt{\frac{1}{1-\gamma_{\mathrm{th}}\left(\kappa_t^2 + \kappa_r^2\right)} \frac{\gamma_{\mathrm{th}}}{\gamma_{s}}} \right),  &  \text{ for } \gamma_{\mathrm{th}} \leq \frac{1}{\kappa_{t}^2 + \kappa_r^2} \\ 1, & \text{ for } \gamma_{\mathrm{th}} > \frac{1}{\kappa_{t}^2 + \kappa_r^2} \end{array}\right..
 	\label{Eq:P_o_THz_s3} 
 \end{align}
\hrulefill
\end{figure*} 
In~\eqref{Eq:P_o_THz_s3}, $\gamma_s$ is the transmission SNR multiplied by the deterministic path-gain, i.e., 
\begin{align}
	\gamma_s = \frac{P_s \prod_{i=1}^{N} g_i}{N_o}.
\end{align}
Note that from~\eqref{Eq:P_o_THz_s3}, it becomes evident that hardware imperfections set a limit to the maximum allowed spectral efficiency of the transmission scheme. In more detail, since the spectral efficiency, $p$, is connected with the SNR threshold through $p=\log_{2}\left(1+\gamma_{th}\right)$, choosing a $p$  greater than $\log_{2}\left(1+\frac{1}{\kappa_t^2+\kappa_r^2}\right)$ would lead to an OP equal to~$1$. 

{The following lemma returns the diversity order of the multi-RIS empowered THz system.
	\begin{lem}
			The diversity order of the multi-RIS empowered THz system can be obtained~as
			\begin{align}
				D_{\text{THZ}} = \min_{i=1,\cdots,2N+L}\mathcal{B}^{\text{THz}}_i/2,
				\label{Eq:D_THZ}
			\end{align}
		where $\mathcal{B}^{\text{THz}}_i$ is the $i-$th element of the tuple 
		\begin{align}
			\mathcal{B}^{\text{THz}}=\left[\alpha_1^{\text{THz}}, \cdots, \alpha_N^{\text{THz}}, \beta_1^{\text{THz}}, \cdots, \beta_N^{\text{THz}}, \xi_1, \cdots, \xi_L, 0\right]
		\end{align}
	\end{lem}
\begin{IEEEproof}
	This lemma can be proven by following the same steps as the proof of Lemma~2. 
\end{IEEEproof}
}

\vspace{-0.3cm}
 \section{Results \& Discussion}\label{S:Results}
 \vspace{-0.2cm}
 
 This section focuses on verifying the theoretical framework presented in Section~\ref{S:SC} by means of  Monte Carlo simulations. To this end, lines are used to denote analytical results, while markers are employed for simulations. The rest of this section is organized as Section~\ref{SS:CMR} focuses on presenting numerical results that quantify the outage performance of cascaded multi-RIS-empowered FSO systems, whereas Section~\ref{SS:PMR} verifies the outage performance bounds of parallel multi-RIS-empowered FSO systems. Finally, in Section~\ref{SS:CRT}, the outage performance of cascaded multi-RIS-empowered THz wireless systems is assessed.    
 
 \vspace{-0.3cm}
 \subsection{Cascaded multi-RIS-empowered FSO systems}\label{SS:CMR} 
 \vspace{-0.1cm}
 
\begin{figure}
	\centering\includegraphics[width=0.67\linewidth,trim=0 0 0 0,clip=false]{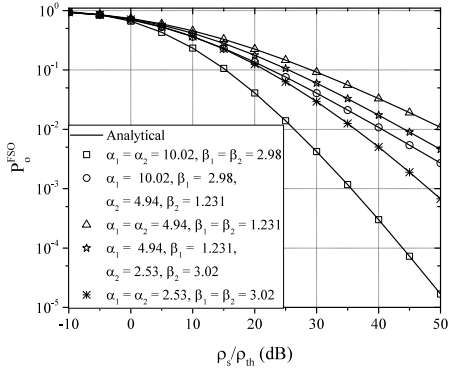}
	\vspace{-0.2cm}
	\caption{OP vs $\rho_{s}/\rho_{\mathrm{th}}$, for different levels of turbulence, assuming $N=2$ and $L=0$.}
	 \vspace{-0.5cm}
	\label{Fig:OP_turbulence}
\end{figure}

Figure~\ref{Fig:OP_turbulence} quantifies the impact of turbulence in a multi-RIS-empowered FSO system, where $N=2$ and $L=0$. In more detail, the OP is plotted against $\rho_{s}/\rho_{\mathrm{th}}$ for different levels of turbulence. The following scenarios are considered: i) both the $S$-RIS and RIS-$D$ links experience weak turbulence, i.e., $\alpha_1=\alpha_2=10.02$ and $\beta_1=\beta_2=2.98$, ii) either the $S$-RIS or the RIS-$D$ link experience weak  ($\alpha_1=10.02$ and $\beta_1=2.98$), while the other suffers from strong turbulence ($\alpha_2=4.942$ and $\beta_1=1.231$), iii) both links experience strong turbulence, iv) one link suffers from strong, while the other from moderate ($\alpha_2=2.53$ and $\beta_1=3.02$) turbulence and v) both links experience from moderate turbulence. This figure reveals that for fixed turbulence conditions, the RIS-empowered FSO system outage performance improves, as $\rho_{s}/\rho_{\mathrm{th}}$ increases. For example, for $\alpha_1=\alpha_2=10.02$ and $\beta_1=\beta_2=2.98$, as $\rho_{s}/\rho_{\mathrm{th}}$ changes from $25$ to $35\text{ }\mathrm{dB}$, the OP decreases by about $10$ times. Additionally, we observe that, for a fixed $\rho_{s}/\rho_{\mathrm{th}}$, as the intensity of turbulence increases in either one of the $S$-RIS and RIS-$D$ links, the outage performance degrades. For instance, for $\rho_{s}/\rho_{\mathrm{th}}=35\text{ }\mathrm{dB}$, the OP increases for more than an order of magnitude as turbulence conditions changes from scenario i to ii. Interestingly, for a fixed $\rho_{s}/\rho_{\mathrm{th}}$, the systems in which both links suffer from moderate turbulence may outperform the ones in which one link suffers from weak and the other from strong turbulence. This highlights the importance of accurately characterizing  turbulence intensity when assessing the performance of RIS-empowered FSO~systems.          
 
 \begin{figure}
 	\centering\includegraphics[width=0.67\linewidth,trim=0 0 0 0,clip=false]{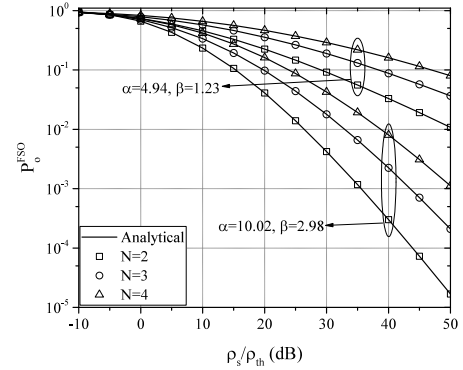}
 	\vspace{-0.2cm}
 	\caption{OP vs $\rho_{s}/\rho_{\mathrm{th}}$, for different values of $N$ and levels of turbulence, assuming $L=0$.}
 	 \vspace{-0.5cm}
 	\label{Fig:OP_N}
 \end{figure}
 
 Figure~\ref{Fig:OP_N} assesses the impact of turbulence in FSO systems that employ multiple RISs. In particular, the OP is depicted as a function of $\rho_{s}/\rho_{\mathrm{th}}$, for different values of $N$ and turbulence, assuming $L=0$. For the turbulence intensity, the following two scenarios are considered: i) the best-case scenario, in which all the links suffer from weak turbulence, i.e., $\alpha_1=\alpha_2=\cdots=\alpha_N=\alpha=10.02$ and $\beta_1=\beta_2=\cdots=\beta_{N}=\beta=2.98$; and ii) the worst-case scenario, in which all the links experience strong turbulence, i.e., $\alpha_1=\alpha_2=\cdots=\alpha_N=\alpha=4.94$ and $\beta_1=\beta_2=\cdots=\beta_{N}=\beta=1.23$. As expected for given $N$ and turbulence conditions, as $\rho_s/\rho_{\mathrm{th}}$ increases, the OP decreases. Moreover, for given $N$ and $\rho_{s}/\rho_{\mathrm{th}}$, as the turbulence intensity gets higher, the OP also increases. For example, for $N=2$ and  $\rho_s/\rho_{\mathrm{th}}=40\text{ }\mathrm{dB}$, the OP increases for about $2$ orders of magnitude, as we move from the best- to the worst-case scenario.  Finally,  for fixed turbulence conditions and $\rho_{s}/\rho_{\mathrm{th}}$, as $N$ increases, the impact of turbulence is multiplied; thus, the OP increases. For instance, for the best-case scenario and $\rho_s/\rho_{\mathrm{th}}=40\text{ }\mathrm{dB}$, the OP increases for about one order of magnitude, as $N$ increases from $2$ to~$3$. This reveals the proliferative effect of turbulence in FSO systems that consists of multiple-RISs.         
 
 \begin{figure}
 	\centering\includegraphics[width=0.67\linewidth,trim=0 0 0 0,clip=false]{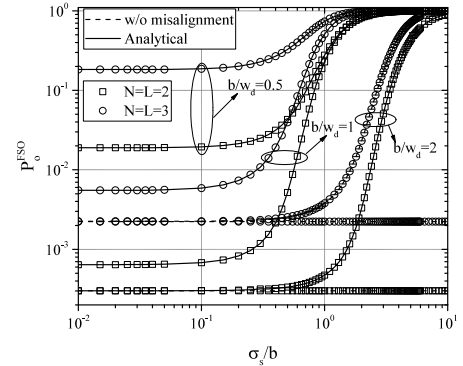}
 	\vspace{-0.2cm}
 	\caption{OP vs $\sigma_{s}/\beta$, for different values of $\beta/w_{d}$ and $N=L$.}
 	 \vspace{-0.5cm}
 	\label{Fig:OP_ss}
 \end{figure}

Figure~\ref{Fig:OP_ss} depicts the OP as a function of the $\sigma_s/b$ for different values of $b/w_d$ and $N=L$, assuming $\rho_s/\rho_{th}=40\text{ }\mathrm{dB}$. Notice that in the scenario under investigation, we assume all the same transmission and misalignment characteristics for all the links. Moreover, the following  scenarios are examined: both RIS's and D lens radius are (i) half, (ii) equal to, and (iii) twice  the radiation beam footprint. As a benchmark, the OP for the case in which there is no misalignment is also presented. As expected, for given $N=L$ and $b/w_{d}$, as $\sigma_s/b$ increases, i.e., the misalignment intensity becomes more severe, the outage performance degrades. For example, for $N=L=2$ and $b/w_{d}=1$, the OP increases for more than one order of magnitude, as $\sigma_s/b$ changes from $0.1$ to $0.5$. Likewise, for fixed $\sigma_s/b$ and $b/w_{d}$, as $N=L$ increases, the joint effect of misalignment and turbulence becomes more severe; thus, the OP increases. Finally, for given $M=N$ and $\sigma_s/b$, as $b/w_{d}$ increases, the beam-waist is constrained; as a result, an outage performance improvement is observed. Interestingly, even in the low-$\sigma_s/b$ regime, if $b/w_d\leq 1$, there exists an OP gap between the two systems that consider and ignore the misalignment fading. This is because~\eqref{Eq:F_Z_1} does not account for the beam-waist. On the other hand, for $b/w_d > 1$, in the low-$\sigma_s/b$ regime, the OP of the FSO system that suffers from misalignment fading tends to the one with misalignment fading immunity. This indicates the importance of taking into account $b/w_d$, when assessing the outage performance of RIS-empowered FSO~systems.

 \begin{figure}
	\centering\includegraphics[width=0.75\linewidth,trim=0 0 0 0,clip=false]{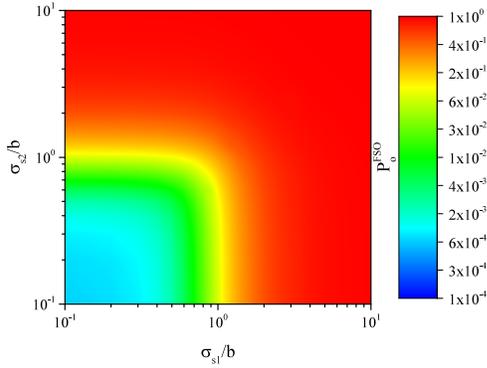}
	\vspace{-0.2cm}
	\caption{OP vs $\sigma_{s1}/\beta$ and $\sigma_{s2}/\beta$, for~$N=L=2$.}
	 \vspace{-0.5cm}
	\label{Fig:OP_ss1_ss2}
\end{figure}

Figure~\ref{Fig:OP_ss1_ss2} demonstrates the impact of different levels of misalignment fading on the OP of a multi-RIS-empowered FSO system with $N=L=2$. Specifically, the OP is presented as a function of $\sigma_{s1}/b$ and $\sigma_{s2}/b$, assuming $\rho_s/\rho_{th}=40\text{ }\mathrm{dB}$, $\alpha_1=\alpha_2=10.02$ and $\beta_1=\beta_2=2.98$. As expected, for a given $\sigma_{s1}/b$, as $\sigma_{s2}/b$ increases, the outage performance degrades. For example, for $\sigma_{s1}/b=0.1$, as $\sigma_{s2}/b$ changes from $0.1$ to $0.6$, the OP increases for approximately one order of magnitude. Similarly, for a fixed $\sigma_{s2}/b$, as $\sigma_{s1}/b$ increases, the OP increases. This indicates that the performance of the multi-RIS-empowered FSO system is determined by the performance of its worst link. Finally, from this figure, we observe that a multi-RIS-empowered FSO system in which $\sigma_1/b=v_1$ and  $\sigma_1/b=v_2$ achieves the same performance as the one with  $\sigma_1/b=v_2$ and~$\sigma_1/b=v_1$.  

\begin{figure}
	\centering\includegraphics[width=0.67\linewidth,trim=0 0 0 0,clip=false]{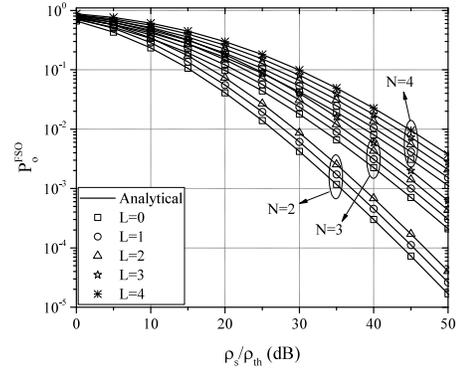}\vspace{-0.2cm}
	\caption{OP vs $\rho_s/\rho_{th}$, for different values of $N$ and $L$.}
	  \vspace{-0.4cm}
	\label{Fig:OP_N_L}
\end{figure}    

Figure~\ref{Fig:OP_N_L} depicts the OP as a function of $\rho_{s}/\rho_{\mathrm{th}}$ for different values of $N$ and $L$, assuming $\alpha_{1}=\alpha_{2}=\cdots=\alpha_{N}=10.02$ and $\beta_{1}=\beta_{2}=\cdots=\beta_{N}=2.98$. As expected, for fixed $N$ and $L$, as $\rho_{s}/\rho_{\mathrm{th}}$ increases, the OP decreases. For instance, for $N=3$ and $L=2$, the outage performance improves for about one order of magnitude, as $\rho_s/\rho_{\mathrm{th}}$ increases from $20$ to $35\text{ }\mathrm{dB}$. Additionally, for given $L$ and $\rho_{s}/\rho_{\mathrm{th}}$, performance degradation is observed as $N$ increases. Finally, for given $N$ and $\rho_{s}/\rho_{\mathrm{th}}$, as $L$ increases, the impact of misalignment becomes more severe; hence, the OP increases.   

 \vspace{-0.3cm}
 \subsection{Parallel Multi-RIS FSO System}\label{SS:PMR}   
  \vspace{-0.1cm}
  
 \begin{figure}
 	\centering\includegraphics[width=0.67\linewidth,trim=0 0 0 0,clip=false]{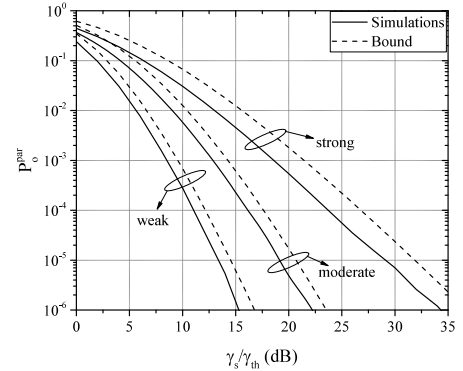}\vspace{-0.2cm}
 	\caption{OP vs $\rho_s/\rho_{th}$, for different levels of turbulence, assuming $N=L=2$.}
 	 \vspace{-0.5cm}
 	\label{Fig:OP_par}
 \end{figure}      
 
 In Fig.~\ref{Fig:OP_par}, the OP of the parallel multi-RIS FSO system is presented as a function of $\rho_s/\rho_{th}$, for different turbulence conditions, assuming $N=L=2$, $\sigma_{s1}/b=\sigma_{s2}/b=0.1$ and $b/w_{d}=0.5$. In this figure, continuous lines are used to denote results extracted from Monte Carlo simulations, while dashed-ones are obtained according to~\eqref{Eq:Po_bound}. It is observed that for a given turbulence condition, as $\rho_{s}/\rho_{\mathrm{th}}$ increases, the OP decreases. Moreover, from this figure, the upper bound is verified. Finally, it becomes apparent that, for a fixed $\rho_{s}/\rho_{\mathrm{th}}$, as the turbulence intensity increases, the error between the simulation and upper bound increases. Despite this fact, the OP upper bound presented in~\eqref{Eq:Po_bound} can become a useful tool for the design of parallel multi-RIS FSO systems.  
 
 \begin{figure}
 	\centering\includegraphics[width=0.67\linewidth,trim=0 0 0 0,clip=false]{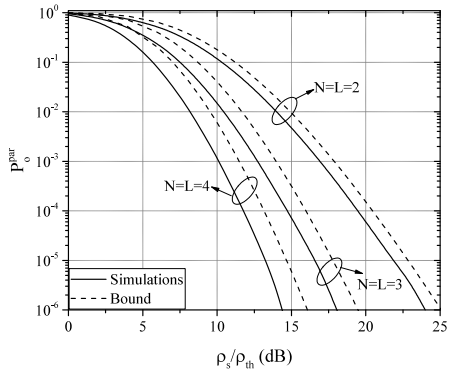}\vspace{-0.2cm}
 	\caption{OP vs $\rho_s/\rho_{th}$, for different values of $N=L$.}
 	 \vspace{-0.5cm}
 	\label{Fig:OP_par_2}
 \end{figure}  

Figure~\ref{Fig:OP_par_2} illustrates the OP of parallel multi-RIS FSO system as a function of $\rho_s/\rho_{th}$, for different values of $N=L$, assuming $\alpha_i=10.02$, $\beta_i=2.98$, $\Omega_i=1$, $\sigma_{si}/b_i=0.1$ and $\beta_i/w_{di}=0.5$ for $i\in[1, N]$. Again,  continuous lines are used to denote results, which are extracted from simulations, while dashed-ones represent the OP upper bound, as derived from~\eqref{Eq:Po_bound}. As expected, for a fixed $N=L$, both the OP and its bound decreases, as $\rho_{s}/\rho_{\mathrm{th}}$ increases. Additionally, it is observed that for a given $\rho_{s}/\rho_{\mathrm{th}}$, as $N=L$ increases, the diversity order increases; thus, outage performance improves. Meanwhile, as $N=L$ increases, the error between the simulations and the OP upper bound increases. Finally, from this figure, it becomes evident that the bound accurately follows the simulations; as a consequence, it can be used to assess the outage performance ceiling of the parallel multi-RIS-empowered FSO system.       
 
  \vspace{-0.3cm}
 \subsection{Cascaded RIS-empowered THz wireless systems} \label{SS:CRT}
  \vspace{-0.1cm}
  
 For the cascaded RIS-empowered THz wireless systems, we consider the following insightful scenario. Unless otherwise stated, $G_s=G_t=50\text{ }\mathrm{dBi}$, $R_i=1$ for $i\in[1, N-1]$ and $f=300\text{ }\mathrm{GHz}$. Additionally, standard atmospheric conditions, i.e., relative humidity, temperature, and pressure  respectively equal to $50\%$, $296\,^{o}K$, and $101325\text{ }\mathrm{Pa}$, are assumed. Under these atmospheric conditions, a realistic value for $C_{n}^{2}$ is $2.3\times 10^{-9}\text{ }\mathrm{m}^{-2/3}$~\cite{Taherkhani2018} and $\kappa(f)=5.8268\times 10^{-4}$.  
 
 \begin{figure}
 	\centering\includegraphics[width=0.75\linewidth,trim=0 0 0 0,clip=false]{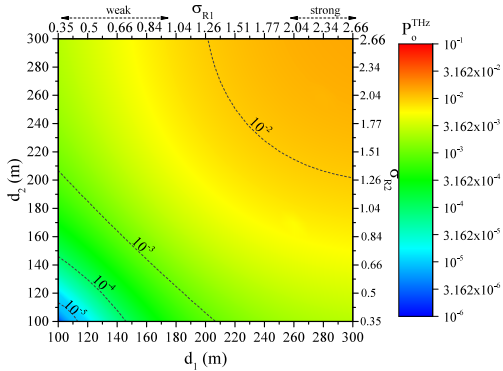}\vspace{-0.2cm}
 	\caption{OP vs $d_1$ and $d_2$.}
 	 \vspace{-0.5cm}
 	\label{Fig:OP_d1_d2}
 \end{figure}

Figure~\ref{Fig:OP_d1_d2} quantifies the impact of turbulence on the outage performance of multi-RIS-empowered THz wireless systems in the absence of misalignment and hardware imperfections, assuming $N=2$, $\gamma_s/\gamma_{\mathrm{th}}=25\text{ }\mathrm{dB}$. In more detail, the OP is illustrated as a function of $d_1$ and $d_2$. In this figure, for the sake of convenience, the corresponding values of $\sigma_{R1}$ and $\sigma_{R2}$ are respectively provided in the top horizontal and right vertical axes. Of note, according to~\cite{Taherkhani2020}, $\sigma_{Ri}<1$,  $1<\sigma_{Ri}<2$ and $\sigma_{Ri}>2$ respectively correspond to weak, moderate, and strong turbulence conditions.  As expected, for a fixed $d_1$, as $d_2$ increases, $\sigma_{R2}$ increases; thus, an outage performance degradation is observed. For example, for $d_{1}=100\text{ }\mathrm{m}$, the OP increases by approximately one order of magnitude, as $d_2$ changes from $110$ to $150\text{ }\mathrm{m}$. Similarly, for a given $d_2$, as $d_1$ increases, $\sigma_{R1}$ also increases, i.e., turbulence intensity increases; in turn, the OP increases. For instance, for $d_{2}=100\text{ }\mathrm{m}$, the OP increases by about $10$ times, as $d_2$ changes from $110$ to $150\text{ }\mathrm{m}$. The aforementioned examples reveal that the system with $d_1=v_1$ and $d_2=v_2$ achieves the same outage performance as the one with $d_1=v_2$ and $d_1=v_1$, where $v_1$ and $v_2$ are independent variables. Finally, from this figure, we observe that for a given e2e transmission distance, $d_1+d_2$, the worst outage performance is observed for $d_1=d_2$. For example, for $d_1+d_2=300\text{ }\mathrm{m}$, the OP for the case in which $d_1=100\text{ }\mathrm{m}$ and $d_2=200\text{ }\mathrm{m}$ is equal to $8.58\times 10^{-4}$, while, for $d_1=d_2=150\text{ }\mathrm{m}$, it is $8.8\times 10^{-4}$. Of note, this comes in line with the majorization theory~\cite{Marshall2010}. 

 \begin{figure}
	\centering\includegraphics[width=0.75\linewidth,trim=0 0 0 0,clip=false]{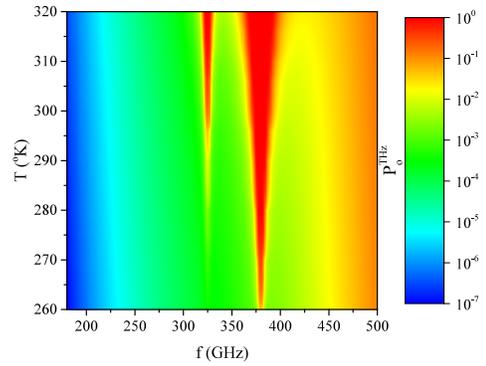}\vspace{-0.2cm}
	\caption{OP vs $f$ and $T$.}
	 \vspace{-0.5cm}
	\label{Fig:OP_T_f}
\end{figure}

{Figure~\ref{Fig:OP_T_f} illustrates the joint effect of molecular absorption and turbulence in terms of OP. In particular, the OP is plotted against the transmission frequency and temperature. The following insightful scenario is considered. The transmission power, bandwidth, $B$, and spectral efficiency of the transmission scheme are respectively set to $0\text{ }\mathrm{dBW}$, $50\text{ }\mathrm{GHz}$, and $2\text{ }\mathrm{bit/s/Hz}$, while $N$ and $L$ are respectively equal to $2$ and $0$. Moreover, $d_1=d_2=100\text{ }\mathrm{m}$ and $\kappa_t=\kappa_r=0$. At the destination side, we assumed that the low noise amplifier's gain and noise figure (NF) are $35\text{ }\mathrm{dB}$, and $1\text{ }\mathrm{dB}$, respectively. The mixer and miscellaneous losses are respectively $5$ and $3\text{ }\mathrm{dB}$, whereas, the mixer's NF is $6\text{ }\mathrm{dB}$. The thermal noise power is evaluated as $N_1 = k_B\, T\,B$, where $k_B$ is the Boltzman's constant. Note that the simulation parameters that we used in this work are inline with~\cite{Boulogeorgos2021b}. This figures shows that two molecular absorption ``walls'' exist around $325$ and $380\text{ }\mathrm{GHz}$. Within the molecular absorption walls, the OP is higher than $10^{-2}$. The length of these walls depends on the temperature. Specifically, as the temperature increases, the molecular absorption becomes more severe; as a result, the length of the molecular absorption walls also increases. Outside the aforementioned walls, in the $100$ to $500\text{ }\mathrm{GHz}$ band, there exist three transmission windows. Within these windows, for a fixed temperature, as the transmission frequency increases, the system outage performance degrades. Similarly, for a given transmission frequency, as the temperature increases, the OP also increases. It is worth noting that, within the transmission windows, the dominant phenomenon that affects the losses is free space propagation, while molecular absorption plays a secondary role.}

\begin{figure}
	\centering\includegraphics[width=0.67\linewidth,trim=0 0 0 0,clip=false]{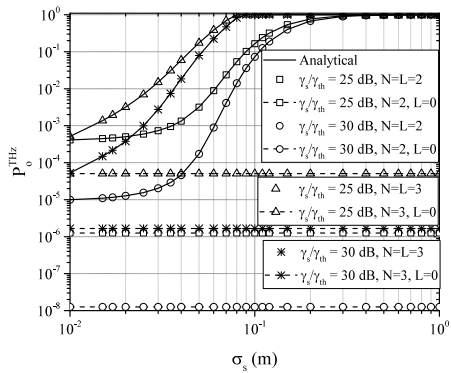}\vspace{-0.2cm}
	\caption{OP vs $\sigma_s$, for different values of $N$, $L$ and $\gamma_s/\gamma_{\mathrm{th}}$.}
	 \vspace{-0.5cm}
	\label{Fig:OP_N_L_gs}
\end{figure}        
 
 Figure~\ref{Fig:OP_N_L_gs} quantifies the outage performance of multi-RIS empowered THz wireless systems in the presence of turbulence and misalignment, assuming that its transceivers are equipped with ideal RF front-end, i.e., $\kappa_t=\kappa_r=0$, and the transmission distances for all the links are set to $100\text{ }\mathrm{m}$. Specifically, the OP is given as a function of $\sigma_s$, for different values of $N=L$ and $\gamma_{s}/\gamma_{\mathrm{th}}$. Of note, in this scenario, we assumed that $\sigma_{s,1}=\sigma_{s,2}=\cdots=\sigma_{s,L}=\sigma_s$. As a benchmark, the OP in the absence of misalignment fading is also plotted. As expected, for given $N=L$ and $\gamma_{s}/\gamma_{\mathrm{th}}$, the outage performance degrades, as the intensity of misalignment fading, i.e., $\sigma_s$, becomes more severe. For example, for $N=L=3$ and $\gamma_{s}/\gamma_{\mathrm{th}}=30\text{ }\mathrm{dB}$, the OP increases for about four orders of magnitude, as $\sigma_s$ changes from $1\text{ }\mathrm{cm}$ to $1\text{ }\mathrm{dm}$. Meanwhile, for fixed $N=L$ and $\sigma_s$, as $\gamma_{s}/\gamma_{\mathrm{th}}$, the OP decreases. On the other hand, for given $\sigma_s$ and $\gamma_{s}/\gamma_{\mathrm{th}}$, as $N=L$ increases, the OP also increases. For instance, for $\sigma_s=4\text{ }\mathrm{cm}$ and $\gamma_{s}/\gamma_{\mathrm{th}}=25\text{ }\mathrm{dB}$, the OP increases by approximately $10$ times, as $N=L$ changes from $2$ to $3$. Finally, from this figure, the detrimental impact of misalignment fading becomes apparent by comparing; thus, the importance of accurately characterizing the channels of the RIS-empowered THz wireless systems is highlighted. 
 
 \begin{figure}
 	\centering\includegraphics[width=0.75\linewidth,trim=0 0 0 0,clip=false]{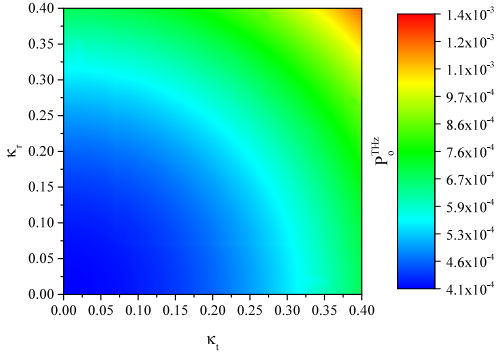}\vspace{-0.2cm}
 	\caption{OP vs $\kappa_t$ and $\kappa_r$.}
 	 \vspace{-0.4cm}
 	\label{Fig:OP_HW}
 \end{figure}

Figure~\ref{Fig:OP_HW} illustrates the impact of hardware imperfections in multi-RIS-empowered THz wireless systems, assuming $L=M=2$, $d_1=d_2=100\text{ }\mathrm{m}$, $\sigma_1=\sigma_2=1\text{ }\mathrm{mm}$ and $\gamma_{s}/\gamma_{th}=25\text{ }\mathrm{dB}$. Of note the case in which $\kappa_t=\kappa_r=0$ corresponds to the best-case scenario, in which both the S and D are equipped with ideal RF front-ends. We observe that, for a given $\kappa_{t}$, as $\kappa_{r}$ increases, the OP also increases. Similarly, for a fixed $\kappa_{t}$, an outage performance degradation occurs, when $\kappa_{r}$ increases. Additionally, for a constant $\kappa_{t}+\kappa_r$, we observe that the OP is maximized for $\kappa_{t}=\kappa_{r}$. Finally,  it is verified that systems with the same $\kappa_{t}^2+\kappa_r^2$ achieve the same outage performance.   

 \begin{figure}
	\centering\includegraphics[width=0.67\linewidth,trim=0 0 0 0,clip=false]{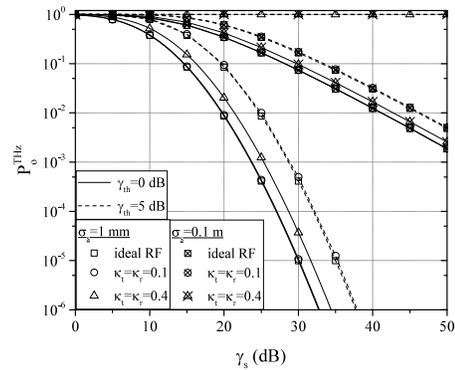}\vspace{-0.2cm}
	\caption{OP vs $\gamma_{s}$, for different values of $\gamma_{\mathrm{th}}$, $\sigma_s$, $\kappa_t$ and $\kappa_r$.}
	 \vspace{-0.5cm}
	\label{Fig:OP_gs_gth}
\end{figure}  

In Fig.~\ref{Fig:OP_gs_gth}, the OP is presented as a function of $\gamma_{s}$ for different values of $\sigma_1=\sigma_2=\sigma_s$, $\gamma_{\mathrm{th}}$ as well as  $\kappa_t$ and $\kappa_r$, assuming $N=L=2$ and $d_1=d_2=100\text{ }\mathrm{m}$. In particular, the ideal ($\kappa_t=\kappa_r=0$), best ($\kappa_t=\kappa_r=0.1$)  and worst ($\kappa_t=\kappa_r=0.4$) practical cases from the hardware imperfections point of view are illustrated. As expected, for given $\gamma_{\mathrm{th}}\leq\frac{1}{\kappa_t^2+\kappa_r^2}$, $\kappa_t$, $\kappa_r$ and $\sigma_{s}$, as $\gamma_s$ increases, the OP decreases. For example, for ideal RF front-end, $\gamma_{\mathrm{th}}=0\text{ }\mathrm{dB}$ and $\sigma_s=1\text{ }\mathrm{mm}$, the OP decreases by $100$ times, as $\gamma_s$ changes from $20$ to $30\text{ }\mathrm{dB}$. On the other hand, as predicted by~\eqref{Eq:P_o_THz_s3}, the OP is equal to $1$. Meanwhile, for fixed $\gamma_{\mathrm{th}}\leq\frac{1}{\kappa_t^2+\kappa_r^2}$, $\kappa_t$, $\kappa_r$ and $\gamma_{s}$, as $\sigma_s$ increases, the OP also increases. For instance, for $\kappa_t=\kappa_r=0.1$, $\gamma_{\mathrm{th}}=0\text{ }\mathrm{dB}$ and $\gamma_s=20\text{ }\mathrm{dB}$, the outage performance degrades by about one order of magnitude, as $\sigma_s$ increases from $1\text{ }\mathrm{mm}$ to $0.1\text{ }\mathrm{m}$. This example highlights the significance of the impact of misalignment fading in multi-RIS-empowered THz wireless systems. Likewise, for given $\sigma_s$, $\kappa_t$, $\kappa_r$ and $\gamma_{s}$, a $\gamma_{\mathrm{th}}$ increase causes an OP increase. Finally, the impact of hardware imperfections in systems operating at higher $\gamma_{\mathrm{th}}$ is more severe than the ones working in lower $\gamma_{\mathrm{th}}$. Since $\gamma_{\mathrm{th}}$ is connected with the transmission scheme's spectral efficiency, the impact of hardware imperfections can be constrained by selecting transmission schemes of low-spectral efficiency. 
 
  \vspace{-0.3cm}
 \section{Conclusions}\label{S:Conclusions}
  \vspace{-0.2cm}
 In this paper, we introduced a theoretical framework for the statistical characterization of cascaded composite turbulence and misalignment channels in terms of PDF and CDF. Moreover, we assessed the outage performance of cascaded multi-RIS FSO and THz systems in the presence of misalignment and turbulence by computing novel closed-form expressions for their OP. Additionally, we provided an insightful upper bound for parallel multi-RIS FSO systems. Our results revealed that in cascaded multi-RIS FSO and THz systems, as the number of RIS and thus the distance between S and D increases, the turbulence and misalignment fading intensity becomes more severe; as a result, the outage performance degrades. On the other hand, in parallel multi-RIS FSO systems, as the number of links increases, the diversity order also increases; hence, the OP decreases. Likewise, the impact of unavoidable transceivers' hardware imperfections on the outage performance of cascaded multi-RIS THz systems was~quantified. Finally, the importance of accurately characterizing the transmission medium was emphasized.        
 
  \vspace{-0.2cm}
  \section*{Appendices}
   \vspace{-0.1cm}
  
\section*{Appendix A}
\vspace{-0.1cm}
\section*{Proof of Theorem 2}
\vspace{-0.1cm}

In order to prove that the PDF of $Z_2$ can be expressed as in~\eqref{Eq:f_Z_2}, we use the induction method~\cite{Gunderson2011}. First, we prove that~\eqref{Eq:f_Z_2} holds for $L=2$. In this case, since $l_1$ and $l_2$ are independent, the PDF of $Z_2$ can be evaluated~as 
\begin{align}
	f_{Z_2}(x;L=2) = \int_{\frac{x}{A_{o,1}}}^{{A_{o,2}}} \frac{1}{y} f_{l_1}\left(\frac{x}{y}\right) f_{l_2}\left(y\right) \, \mathrm{d}y.
	\label{Eq:f_Z_2_L_2_s1}
\end{align}
By applying~\eqref{Eq:f_l_i} in~\eqref{Eq:f_Z_2_L_2_s1}, we obtain
\begin{align}
	f_{Z_2}(x;L=2) = \frac{\xi_1 \xi_2}{A_{o,1}^{\xi_1} A_{o,2}^{\xi_2}} x^{\xi_1-1}\int_{\frac{x}{A_{o,1}}}^{{A_{o,2}}}y^{-1}\,\mathrm{d}y,
\end{align}
which, according to~\cite[eq. (2.01/2)]{B:Gra_Ryz_Book}, can be rewritten~as
\begin{align}
	f_{Z_2}(x;L=2) = \frac{\prod_{i=1}^{2}\xi_i}{\prod_{i=1}^{2}A_{o,i}^{\xi_i}} x^{\xi_1-1}\ln\left(\frac{\prod_{i=1}^{2}A_{o,i}}{x}\right), 
	\label{Eq:f_Z_2_L_2_final}
\end{align}
for  $0\leq x \leq A_{o,1} A_{o,2}$. 
By comparing~\eqref{Eq:f_Z_2_L_2_final} with~\eqref{Eq:f_Z_2}, it becomes evident that~\eqref{Eq:f_Z_2} is true for $L=2$. 

Next, for $L=3$, since $l_1$, $l_2$, and $l_3$ are independent, the PDF of $Z_2$ can be obtained~as
\begin{align}
f_{Z_2}(x;L=3) = \int_{\frac{x}{A_{o,3}}}^{{\prod_{i=1}^{2}A_{o,i}}} \frac{1}{y}\, f_{Z_2}\left(y;L=2\right)\, f_{l_3}\left(\frac{x}{y}\right)\,\mathrm{d}y,
\end{align}
which, with the aid of~\eqref{Eq:f_l_i} and~\eqref{Eq:f_Z_2_L_2_final}, can be rewritten~as
\begin{align}
	f_{Z_2}(x;L=3) &= \prod_{i=1}^{3}\frac{\xi_i}{A_{o,i}^{\xi_i}} x^{\xi_3-1}
	\nonumber \\ & \times
	\int_{\frac{x}{A_{o,3}}}^{{\prod_{i=1}^{2}A_{o,i}}} y^{-1}\ln\left(\frac{\prod_{i=1}^{2}A_{o,i}}{y}\right)\,\mathrm{d}y.
	\label{Eq:f_Z_2_L_3_s_2} 
\end{align}
Next, by applying~\cite[eq. (2.721/3)]{B:Gra_Ryz_Book} in~\eqref{Eq:f_Z_2_L_3_s_2}, we get
\begin{align}
	f_{Z_2}(x;L=3)\hspace{-0.1cm} =\hspace{-0.1cm} \frac{1}{2} \frac{\prod_{i=1}^{3}\xi_i x^{\xi-1}}{\prod_{i=1}^{3}A_{o,i}^{\xi_i}}  \left(\ln\left(\frac{\prod_{i=1}^{3}A_{o,i}^3}{x} \right)\right)^{2},
\end{align}
with $0\leq x \leq A_o^{3}$. 

Let us assume that~\eqref{Eq:f_Z_2} holds for $L=M$, i.e.,
\begin{align}
		f_{Z_{2}}(x;L=M) &= \frac{1}{(M-1)!}
		\frac{\prod_{i=1}^{M}\xi_{i}}{\prod_{i=1}^{M}A_{o,i}^{\xi_i}} x^{\xi_M-1}
		\nonumber \\ & \times \left(\ln\left(\frac{\prod_{i=1}^{M}A_{o,i}}{x}\right)\right)^{M-1}.
		\label{Eq:f_Z_2_L_M}
\end{align}
Then, for $L=M+1$, the PDF of $Z_2$ can be evaluated~as
\begin{align}
	f_{Z_2}(x;L=M+1) = \int_{\frac{x}{A_o}}^{{A_o^{M}}} \frac{f_{Z_2}\left(y;L=M\right)\, f_{l_{M+1}}\left(\frac{x}{y}\right)}{y}\,\mathrm{d}y,
\end{align}
which, by applying~\eqref{Eq:f_l_i} and~\eqref{Eq:f_Z_2_L_M}, can be rewritten~as
\begin{align}
	f_{Z_2}(x;L=M+1) &= \frac{x^{\xi-1}}{(M-1)!}\left(\frac{\xi}{A_o^{\xi}}\right)^{M+1} 
	\nonumber \\ & \times \int_{\frac{x}{A_o}}^{{A_o^{M}}}  \left(\ln\left(\frac{A_o^{M}}{y}\right)\right)^{M-1} \,\mathrm{d}y.
	\label{Eq:f_Z_2_L_M_1}
\end{align}
By applying~\cite[eq. (2.721/1)]{B:Gra_Ryz_Book},~\eqref{Eq:f_Z_2_L_M_1} can be written in a closed-form as
\begin{align}
	f_{Z_2}(x;L=M+1) &= \frac{1}{M!} \frac{\prod_{i=1}^{M+1}\xi_i}{\prod_{i=1}^{M+1}A_{o,i}^{\xi_i}} x^{\xi_{M+1}-1}
	\nonumber \\ & \times \left(\ln\left(\frac{\prod_{i=1}^{M}A_{o,i}}{x}\right)\right)^{M},
	\label{Eq:f_Z_2_L_M_1_final}
\end{align}
with $0\leq x\leq x^{M+1}$. Notice, that from~\eqref{Eq:f_Z_2_L_M_1_final}, it becomes apparent that if~\eqref{Eq:f_Z_2} holds for $L=M$, it also holds for $L=M+1$. Since it also holds for $N-2$, then it stands for each $N\geq 2$. This concludes the proof. 
\vspace{-0.2cm}
\section*{Appendix B}\vspace{-0.2cm}
\section*{Proof of Theorem 3}\vspace{-0.2cm}
Since $L\leq N$,with the aid of~\eqref{Eq:Z_1} and~\eqref{Eq:Z_2},~\eqref{Eq:Z} can be rewritten~as
\begin{align}
	Z = Y_1 \, Y_2,
\end{align}
where
\begin{align}
	Y_1 = \prod_{i=1}^{L} r_i l_i
\quad \text{ and } \quad
	Y_2 = \prod_{i=L+1}^{N} r_i.
\end{align}

To extract a closed-form expression for the PDF of $Z$, we need first to evaluate the PDF of $Y_1$. In this direction, let us assume that $L=1$; then, the PDF of $Y_1$ yields~as
\begin{align}
	f_{Y_1}(x;L=1) = \int_{\frac{x}{A_{o,1}}}^{\infty}\frac{1}{y} f_{r_{1}}\left(y\right)\, f_{l_1}\left(\frac{x}{y}\right)\,\mathrm{d}x,
\end{align}  
which, by applying~\eqref{Eq:f_r_i} and~\eqref{Eq:f_l_i}, can be rewritten~as
\begin{align}
	f_{Y_1}(x;L=1) &= 2\left(\frac{\alpha_1 \beta_1}{\Omega_1}\right)^{\frac{\alpha_1+\beta_1}{2}} \frac{1}{\Gamma\left(\alpha_1\right) \Gamma\left(\beta_1\right)} \frac{\xi}{A_{o,1}^{\xi_1}} x^{\xi_1-1} 
	\nonumber \\ & \hspace{-1.2cm} \times
	\int_{\frac{x}{A}}^{\infty} y^{\frac{\alpha_1+\beta_1}{2}-1-\xi_1} \, \mathrm{K}_{\alpha_1-\beta_1}\left(2\sqrt{\frac{\alpha_1 \beta_1 y}{\Omega_1}}\right)\,\mathrm{d}y.
	\label{Eq:f_Y_1_L_1}
\end{align}
By employing~\cite[ch. 2.6]{Mathai1973},~\eqref{Eq:f_Y_1_L_1} can be written~as in~\eqref{Eq:f_Y_1_L_1_s2}, given at the top of the next page.
\begin{figure*}
\begin{align}
	&f_{Y_1}(x;L=1) = \left(\frac{\alpha_1 \beta_1}{\Omega_1}\right)^{\frac{\alpha_1+\beta_1}{2}} \frac{1}{\Gamma\left(\alpha_1\right) \Gamma\left(\beta_1\right)} \frac{\xi}{A_{o,1}^{\xi_1}} x^{\xi_1-1} 
	\int_{\frac{x}{A_{o,1}}}^{\infty} \left( \frac{1}{y}\right)^{-\frac{\alpha_1+\beta_1}{2}+\xi_1+1} \mathrm{G}_{0, 2}^{2, 0}\left( \frac{\alpha_1 \beta_1 y}{\Omega_1}\left| \frac{\alpha_1-\beta_1}{2}, -\frac{\alpha_1-\beta_1}{2}  \right.\right)\,\mathrm{d}y
	\label{Eq:f_Y_1_L_1_s2}
\end{align}
\hrulefill
\end{figure*}
Additionally, by applying~\cite[eq. (2.24.5/3)]{B:Prudnikov_v3},~\eqref{Eq:f_Y_1_L_1_s2} can be analytically expressed~as
\begin{align}
	&f_{Y_1}(x;L=1) = \left(\frac{\alpha_1 \beta_1 }{\Omega_1 A_{o,1}}\right)^{\frac{\alpha_1+\beta_1}{2}} \frac{\xi_1}{\Gamma\left(\alpha_1\right) \Gamma\left(\beta_1\right)}  x^{\frac{\alpha_1+\beta_1}{2}-1} 
	\nonumber \\ & \times
	\mathrm{G}_{1, 3}^{3, 0}\left(\frac{\alpha_1\beta_1 x}{A_o \Omega_1}\left| \begin{array}{c} -\frac{\alpha_1+\beta_1}{2}+\xi_1+1 \\ \frac{\alpha_1-\beta_1}{2}, -\frac{\alpha_1-\beta_1}{2}, -\frac{\alpha_1+\beta_1}{2}+\xi_1  \end{array}\right. \right), 
\end{align}
which, with the aid of~\cite{WS:mathematica_function}, can be rewritten~as
\begin{align}
	&f_{Y_1}(x;L=1) = \frac{\xi_1 x^{-1}}{\Gamma\left(\alpha_1\right) \Gamma\left(\beta_1\right)}\mathrm{G}_{1,3}^{3, 0}\left(\frac{\alpha_1\beta_1 x}{A_{o,1} \Omega_1}\left|\begin{array}{c} \xi_1+1 \\ \alpha_1, \beta_1, \xi_1 \end{array}\right. \right).
	\label{Eq:f_Y_1_L_1_final}
\end{align}

For $L=2$, the PDF of $Y_1$ can be obtained~as
\begin{align}
	&f_{Y_1}(x;L=2) = \int_{0}^{\infty} \frac{1}{y} f_{Y_1}(y;L=1) \, f_{Y_1}\left(\frac{x}{y};L=1\right) \, \mathrm{d}y,
\end{align}
which, by applying~\eqref{Eq:f_Y_1_L_1_final}, can be rewritten as in~\eqref{Eq:f_Y_1_L_2_s1}, given at the top of the next page.
\begin{figure*}
\begin{align}
	&f_{Y_1}(x;L=2) = \frac{\prod_{i=1}^{2}\xi_i}{\prod_{i=1}^2\Gamma\left(\alpha_i\right) \Gamma\left(\beta_i\right)} x^{-1}
	\int_{0}^{\infty} y^{-1} \mathrm{G}_{1,3}^{3, 0}\left(\frac{\alpha_1\beta_1 y}{A_{o,1} \Omega_1}\left|\begin{array}{c} \xi_1+1 \\ \alpha_1, \beta_1, \xi_1 \end{array}\right. \right) \mathrm{G}_{1,3}^{3, 0}\left(\frac{\alpha_2\beta_2 x}{A_{o,2} \Omega_2 y}\left|\begin{array}{c} \xi_2+1 \\ \alpha_2, \beta_2, \xi_2 \end{array}\right. \right)\, \mathrm{d}y
	\label{Eq:f_Y_1_L_2_s1}
\end{align}
\hrulefill
\end{figure*}
By employing~\cite[eq. (9.31/1)]{B:Gra_Ryz_Book}, we can write~\eqref{Eq:f_Y_1_L_2_s1} as in~\eqref{Eq:f_Y_1_L_2_s2}, given at the top of the next page. 
\begin{figure*}
\begin{align}
	f_{Y_1}(x;L=2) = \frac{\prod_{i=1}^{2}\xi_i}{\prod_{i=1}^2\Gamma\left(\alpha_i\right) \Gamma\left(\beta_i\right)} x^{-1}
	\int_{0}^{\infty} y^{-1} \mathrm{G}_{1,3}^{3, 0}\left(\frac{\alpha_1\beta_1 y}{A_{o,1} \Omega_1}\left|\begin{array}{c} \xi_1+1 \\ \alpha_1, \beta_1, \xi_1 \end{array}\right. \right) 
	\mathrm{G}_{3,1}^{0, 3}\left(\frac{A_{o,2} \Omega_2 y}{\alpha_2\beta_2 x}\left|\begin{array}{c} 1-\alpha_2, 1-\beta_2, 1-\xi_2  \\ 1-\xi_2  \end{array}\right. \right)\, \mathrm{d}y
	\label{Eq:f_Y_1_L_2_s2}
\end{align} 
\hrulefill
\end{figure*}
Finally, by applying~\cite[eq. (2.24.5/3)]{B:Prudnikov_v3} in~\eqref{Eq:f_Y_1_L_2_s2}, we~get
\begin{align}
	f_{Y_1}(x;L=2) &= \frac{\prod_{i=1}^{2}\xi_i}{\prod_{i=1}^2\Gamma\left(\alpha_i\right) \Gamma\left(\beta_i\right)} x^{-1} 
	\nonumber \\ & \hspace{-1.7cm}\times
	\mathrm{G}_{2, 6}^{6, 0}\left(\frac{\prod_{i=1}^{2}\alpha_i\beta_i}{A_{o,1}\prod_{i=1}^{2}\Omega_i}y\left| \begin{array}{c} \xi_1+1, \xi_2+1 \\ \alpha_1, \beta_1, \xi_1, \alpha_2, \beta_2, \xi_2 \end{array}\right. \right).
\label{Eq:f_Y_1_L_2_s3}
\end{align} 	
By recurrently conducting this procedure, for $L=3, 4, \cdots$, we prove that $f_{Y_1}(x)$ can be obtained as in~\eqref{Eq:f_Y_1}, given at the top of the next page.
\begin{figure*}
\begin{align}
	f_{Y_1}(x) &= \frac{\prod_{i=1}^{L}\xi_i}{\prod_{i=1}^L\Gamma\left(\alpha_i\right) \Gamma\left(\beta_i\right)} x^{-1} 
	\mathrm{G}_{L, 3L}^{3L, 0}\left(\frac{\prod_{i=1}^{L}\alpha_i\beta_i}{\prod_{i=1}^{L}\Omega_i A_{o,i}}x\left| \begin{array}{c} \xi_1+1, \xi_2+1, \cdots, \xi_L+1 \\ \alpha_1, \cdots, \alpha_L, \beta_1, \cdots, \alpha_L, \xi_1, \xi_2, \cdots, \xi_L \end{array}\right. \right).
	\label{Eq:f_Y_1}
\end{align} 
\hrulefill
\end{figure*}

Moreover, by following the same steps as in Appendix B, it can be proven that the PDF of $Y_2$ can be obtained as in~\eqref{Eq:f_Y_2}, given at the top of the next page.
\begin{figure*}
 \begin{align}
 	&f_{Y_{2}}(x) = \frac{\mathrm{G}_{0, 2\left(N-L\right)}^{2\left(N-L\right), 0}\left[ x \prod_{i=L+1}^{N} \frac{\alpha_i \beta_i}{\Omega_i} \Big| \alpha_{L+1}, \beta_{L+1}, \alpha_{L+2}, \beta_{L+2} \cdots, \alpha_{N}, \beta_{N} \big. \right]}{x\prod_{i=L+1}^{N} \Gamma\left(\alpha_i\right) \Gamma\left(\beta_i\right)}.
 	\label{Eq:f_Y_2} 
 \end{align}
\hrulefill
\end{figure*} 
Thus, since $Y_1$ and $Y_2$ are independent, the PDF of $Z$ occurs~as
\begin{align}
	f_{Z}(x) = \int_{0}^{\infty} \frac{1}{y} f_{Y_{1}}(y) \, f_{Y_{2}}\left(\frac{x}{y}\right)\, \mathrm{d}y,
\end{align}
which by substituting~\eqref{Eq:f_Y_1} and~\eqref{Eq:f_Y_2}, can be written as in~\eqref{Eq:f_Z_s1}, given at the top of the next page.
\begin{figure*}
\begin{align}
	f_{Z}(x) = \frac{\prod_{i=1}^{L}\xi_i}{\prod_{i=1}^N\Gamma\left(\alpha_i\right) \Gamma\left(\beta_i\right)} 
	x^{-1}
	&\int_{0}^{\infty} y^{-1} 
	\mathrm{G}_{L, 3L}^{3L, 0}\left(\frac{\prod_{i=1}^{L}\alpha_i\beta_i}{\prod_{i=1}^{L}\Omega_i A_{o,i}}y\left| \begin{array}{c} \xi_1+1, \xi_2+1, \cdots, \xi_L+1 \\ \alpha_1, \cdots, \alpha_L, \beta_1, \cdots, \alpha_L, \xi_1, \xi_2, \cdots, \xi_L \end{array}\right. \right)
	\nonumber \\ & \times
	\mathrm{G}_{0, 2\left(N-L\right)}^{2\left(N-L\right), 0}\left( \frac{x}{y} \prod_{i=L+1}^{N} \frac{\alpha_i \beta_i}{\Omega_i} \Big| \alpha_{L+1}, \beta_{L+1}, \alpha_{L+2}, \beta_{L+2} \cdots, \alpha_{N}, \beta_{N} \big. \right) \,\mathrm{d}y
	\label{Eq:f_Z_s1}
\end{align}
\hrulefill
\end{figure*}
By employing~\cite{WS:mathematica_function},~\eqref{Eq:f_Z_s1} can be rewritten in a closed-form as in~\eqref{Eq:f_Z_s2}, given at the top of the next page.
\begin{figure*}
	\begin{align}
		f_{Z}(x) &= \frac{\xi^L}{\prod_{i=1}^N\Gamma\left(\alpha_i\right) \Gamma\left(\beta_i\right)} 
		x^{-1}
		\int_{0}^{\infty} y^{-1} 
		\mathrm{G}_{L, 3L}^{3L, 0}\left(\frac{\prod_{i=1}^{L}\alpha_i\beta_i}{\prod_{i=1}^{L}\Omega_i A_{o,i}}y\left| \begin{array}{c} \xi_1+1, \xi_2+1, \cdots, \xi_L+1 \\ \alpha_1, \cdots, \alpha_L, \beta_1, \cdots, \alpha_L, \xi_1, \xi_2, \cdots, \xi_L \end{array}\right. \right)
		\nonumber \\ & \times
		\mathrm{G}_{2\left(N-L\right),0}^{0,2\left(N-L\right),}\left( \frac{y}{x} \prod_{i=L+1}^{N} \frac{\Omega_i}{\alpha_i \beta_i} \Big| 1-\alpha_{L+1}, 1-\beta_{L+1}, 1-\alpha_{L+2}, 1-\beta_{L+2} \cdots, 1-\alpha_{N}, 1-\beta_{N} \big. \right) \,\mathrm{d}y
		\label{Eq:f_Z_s2}
	\end{align}
	\hrulefill
\end{figure*}
Finally, by employing~\cite[eq. (2.24.5/3)]{B:Prudnikov_v3}, we get~\eqref{Eq:f_Z_final}. 

The CDF of $Z$ can be evaluated~as
\begin{align}
	F_{Z}(x) = \int_{0}^{x} f_{Z}(y)\,\mathrm{d}y,
\end{align} 
which, by applying~\cite[eq. (2.24.5/3)]{B:Prudnikov_v3}, returns~\eqref{Eq:F_Z_final}. This concludes the proof. 

\vspace{-0.5cm}

\balance
\bibliographystyle{IEEEtran}
\bibliography{IEEEabrv,References}

\end{document}